\documentclass[lettersize,journal]{IEEEtran}

\usepackage{cite, makecell}
\usepackage{amsmath,amssymb,amsfonts}
\usepackage{algorithmic}
\usepackage{graphicx}
\usepackage{textcomp}
\usepackage{xcolor, rotating}
\usepackage{balance}
\usepackage{multirow}
\usepackage{gensymb}
\usepackage{array,booktabs}
\usepackage{longtable,tabularx,tabulary}
\usepackage{lipsum}
\usepackage{siunitx}
\usepackage{enumitem}
\usepackage{amssymb,amsmath,epsfig,latexsym, bm}
\usepackage{hyperref}
\hypersetup{colorlinks, pdfa=true, breaklinks, citecolor=blue, urlcolor=black, linkcolor=black}
\usepackage{hyphenat}
\usepackage{subcaption}

\usepackage{pgfplots}
\DeclareUnicodeCharacter{2212}{−}
\usepgfplotslibrary{groupplots,dateplot}
\usetikzlibrary{patterns,shapes.arrows}
\pgfplotsset{compat=newest}
\pgfplotsset{compat=1.11,
    /pgfplots/ybar legend/.style={
    /pgfplots/legend image code/.code={%
       \draw[##1,/tikz/.cd,yshift=-0.25em]
        (0cm,0cm) rectangle (15pt,0.8em);},
   },
}
\ifCLASSINFOpdf
\else
\fi
\begin{document}
\title{Quantum-Hybrid Support Vector Machines for Anomaly Detection in Industrial Control Systems}

\author{Tyler~Cultice,~\IEEEmembership{Student Member, IEEE},
        Md.~Saif~Hassan~Onim,~~\IEEEmembership{Student Member, IEEE},
        Annarita~Giani,~~\IEEEmembership{Senior Member, IEEE},
        and~Himanshu~Thapliyal,~~\IEEEmembership{Senior Member, IEEE}% <-this % stops a space
\thanks{Tyler C., Md. Saif O., and Himanshu T. are with the Department
of Electrical Engineering and Compute Science, University of Tennessee, Knoxville,
TN, 37996 USA. Annarita G. is with GE Vernova Research Center. Contact e-mail: hthapliyal@utk.edu.}% <-this % stops a space
%\thanks{Manuscript received xxxx; revised xxxx.}
}


\markboth{IEEE Transactions on Emerging Topics in Computing}%
{IEEE Transactions on Emerging Topics in Computing}



\maketitle

\begin{abstract}
Sensitive data captured by Industrial Control Systems (ICS) play a large role in the safety and integrity of many critical infrastructures. Detection of anomalous or malicious data, or Anomaly Detection (AD), with machine learning is one of many vital components of cyberphysical security. Quantum kernel-based machine learning methods have shown promise in identifying complex anomalous behavior by leveraging the highly expressive and efficient feature spaces of quantum computing. This study focuses on the parameterization of Quantum Hybrid Support Vector Machines (QSVMs) using three popular datasets from Cyber-Physical Systems (CPS).  The results demonstrate that QSVMs outperform traditional classical kernel methods, achieving 13.3\% higher F1 scores. Additionally, this research investigates noise using simulations based on real IBMQ hardware, revealing a maximum error of only 0.98\% in the QSVM kernels. This error results in an average reduction of 1.57\% in classification metrics. Furthermore, the study found that QSVMs show a 91.023\% improvement in kernel-target alignment compared to classical methods, indicating a potential "quantum advantage" in anomaly detection for critical infrastructures.
This effort suggests that QSVMs can provide a substantial advantage in anomaly detection for ICS, ultimately enhancing the security and integrity of critical infrastructures.
\end{abstract}


\IEEEpeerreviewmaketitle

\section{Introduction} \label{sec:Introduction}
\IEEEPARstart{C}{yberphysical} Systems, also known as CPS, remain one of the critical pillars of many of the fastest-growing cybernetic domains in recent times, especially with the inclusion of Industry 2.0 ``smart" systems. In 2025, it is expected that the CPS market will grow from \$124.1 billion to 255.3 billion from 2024 to 2029~\cite{markets2024}. Industrial Control Systems (ICS), a type of cyberphysical system, brilliantly use sensor-based data and tightly monitored feedback loops controlled by computing to keep industrial infrastructures operational and highly efficient. The integration of Industry 2.0 concepts, such as remote monitoring and data-driven automation, has become a cornerstone of modern operations in industries like power, water treatment, and manufacturing.

The use of cyberattacks to target ICS infrastructure has been of great concern for security experts. A recent analysis revealed a substantial volume of vulnerabilities in CPS, with an average of 115 published vulnerabilities per month in the second half of 2022, with over 82\% of organizations with ICS systems reporting that they have suffered cyber-attacks targetting their IoT-enabled systems~\cite{claroty2022}. Some of these vulnerabilities lead malicious parties to a very critical target: the data.

Anomaly Detection, or AD, in data identifies when sensor data and logs contain anomalous samples from malfunctioning devices, likely victims of malicious interference. Most ICS systems utilize the Supervisory Control and Data Acquisition (SCADA) architecture, which contain monitoring alarms based on rules built-in or user-created. However, many researchers have previously determined that SCADA alarms alone are inadequate for modern security threats~\cite{Cardenas2021}. Thus, more sophisticated methods for AD monitoring have been utilized in the past decade, including machine learning (ML) models to dynamically classify anomalous data and behaviors. However, just as modern ICS systems continue to rapidly grow in complexity and connectivity, the number of attack vectors and potential targets expand as well.

Quantum Machine Learning (QML) can be a pioneer solution to this ever-growing control cybersecurity problem. Using the highly expressive Hilbert spaces of quantum kernels, Quantum-Hybrid Support Vector Machines (QSVMs) can efficiently calculate the dissimilarities of anomalies in real-world data. However, there are many deciding factors and parameters to training an effective diagnostic model. An in-depth analysis and conceptualization using modern, updated quantum systems is mandatory to identify any quantifiable quantum advantage over traditional kernel functions.

This study investigates the application of Quantum Hybrid Support Vector Machines as an anomaly detection technique for identifying unusual patterns in real-world datasets from ICS and Cyber-Physical Systems CPS. By pairing deeply parameterized, state-of-the-art feature maps and unique pre-processing methods, this work aims to thoroughly bridge the gap between theory and practice and identify an advantage of quantum kernels in the modern NISQ era. Our QSVM analysis is conducted using three independent ICS datasets from critical water infrastructure, specifically: (1) water treatment, (2) water distribution, and (3) hydropower generation. 

A quantum advantage is demonstrated, with quantum computing showing improvements over classical computing, highlighting the  benefits of quantum capabilities. This work is an extension of our work presented in \cite{Cultice2024}.
Our contributions can be described as follows:
\begin{itemize}
    \item Implementation of AD-focused QSVM designs using multiple highly effective feature maps from existing literature with an average classification F1 Score of 0.9. 
    \item Investigation into quantum kernel performance using three popular CPS datasets with up to 24 features and 2000 windowed and processed data points each, larger than previous literature today.
    \item Comparison of the best performing quantum models to traditional kernel methods, resulting in an average 13.3\% improvement in F1 score over even the best classical kernel solutions across all three datasets.
    \item Identified a clear quantum advantage of using quantum feature maps by observing the highly accepted geometric kernel distance metric and more.
\end{itemize}

The organization is as follows: Section \ref{sec:Background} provides relevant background on topics in anomaly detection in CPS security and QSVMs. Section \ref{sec:Datasets} provides a breakdown of the three CPS datasets used in this work. Section \ref{sec:Methodology} discusses the framework and methodology. Section \ref{sec:Results} discusses the classification results of the QSVM. Section \ref{sec:Results2} investigates quantum noise, and ``quantum advantage". Section \ref{sec:Conclusion} draws conclusions on the QSVM model and future work.

\section{Background} \label{sec:Background}
\subsection{Classical Anomaly Detection}
Cyber-related vulnerabilities and attack models in critical infrastructures have been surveyed in existing literature~\cite{Ding2021,Chen2019}. Many of these works have shown that small alterations of the data from cyberattacks can result in catastrophic effects on CPS feedback loops~\cite{Talukder2021}.

Deep-convolution neural network models have previously been applied to solve identify complex CPS anomalies~\cite{Cultice2020,Yin2022,Raman2022}. Kim et al.~\cite{Kim2023} investigated the feature importance characteristics of CPS datasets, such as SWaT and HAI. Typically, OCSVMs are well known for being strong anomaly detectors when applied. However, recent literature claim classical SVMs to be weak models for these CPS datasets~\cite{Tushkanova2023}.

\subsection{Embedding Classical Data to Quantum}
In quantum computing, information is represented in a superposition state of both binary states, 0 and 1. This superposition state allows us to effectively represent any value with one qubit. To represent classical values, encoding circuits map data features on each qubit, resulting in a quantum state representing all $N$ features. Angular encoding by far remains the most popular method, which utilizes parametrized rotation gates to encode data into qubit states. Furthermore, dense-angle encoding represents two features per qubit by taking advantage of the relative phase degree of freedom~\cite{LaRose2020}, resulting in $N$ features expressed by $n = N/2$ qubits. We generalized this operation as a unitary $S_x$, which maps the feature vector $x = [x_1,...,x_N]^T \in \mathbb{R}^N$ to $N/2$ qubits.

% Compute Uncompute Figure
\begin{figure}[htbp]
\centering
\includegraphics[width=\columnwidth]{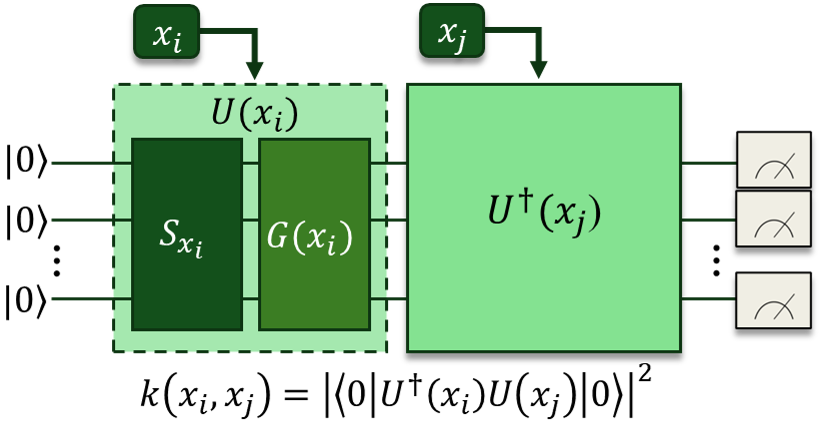}
\caption{Fidelity quantum kernel, $k({x}_{i},{x}_{j})={\left\vert \langle 0| {U}^{{{\dagger}} }({x}_{i})U({x}_{j})| 0\rangle \right\vert }^{2}$. $x_i$ and $x_j$ denote $N$-sized feature vectors of data $x$. A unitary transformation $U(x_i)$ is performed, followed by its inverse $U^\dag(x_j)$ and the measurement. $S_{x_i}$ refers to the encoding circuit, while $G(x_i)$ refers to any interaction.}
\label{fig:computeUncompute}
\end{figure}

\subsection{Quantum Fidelity Kernels}
Quantum-Hybrid Support Vector Machines utilize projected quantum kernel functions to transform the information into a higher dimensional space for a classical SVM. Quantum kernel functions derive from the state fidelity, or overlap, between two quantum states produced by the parameterized ``feature map", $U(x)$. This state fidelity is defined as $K(x_i,x_j) = |\langle\phi(x_i)|\phi(x_j)\rangle|^2$, which is effectively an internal dot product between the image of two vectors prepared as $\phi(x)$. Quantum feature maps~\cite{Schuld2019} encode each feature to a higher-dimensional $n$-qubit Hilbert Space $\mathcal{H}$, resulting in a separable, dimensionally flattened state. Quantum entanglements, defined as $G(x)$, promote interaction in the encoded state.

The most efficient method of computing state fidelity uses the Hilbert-Schmidt inner product, as described by Havlíček et al.~\cite{Havlicek2019}. A visual representation is included in Figure \ref{fig:computeUncompute}. This process applies the feature map unitary $U(x_i)$ followed by its inverse unitary $U^\dag(x_j)$ and a measurement. The state fidelity or overlap of $x_i$ and $x_j$ directly characterizes the similarity between the two data points in $x$.  When this fidelity operation is repeated for all pairs of $x_i$ and $x_j$, the result is an inner-product matrix of all fidelities bounded by $K(x_i,x_j) \rightarrow [0,1]$.

\begin{figure*}[htb]
    \centering
    \includegraphics[width=1.9\columnwidth]{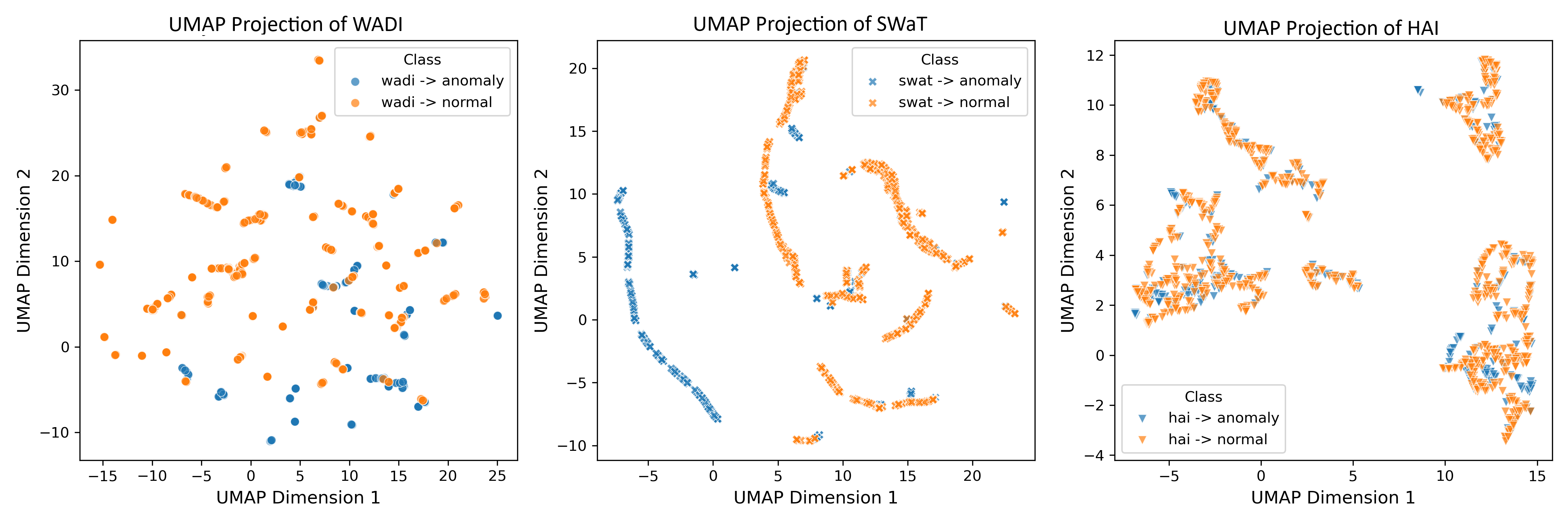}
    \caption{UMAP Projection of Class Distribution in the Test Data for the SWaT, HAI, and WADI datasets.}
    \label{fig:proj}
\end{figure*}

\subsection{Kernel Alignment as a Similarity Metric} \label{sec:kernelAlign}
The best way to quantify the similarities between two kernels is to calculate the inner product between two kernels using Kernel Alignment~\cite{Cristianini2001}, defined as

\begin{equation}
\label{eq:polarity}
    \operatorname{KA}(K_1, K_2) = \frac{\operatorname{Tr}(K_1 K_2)}{\sqrt{\operatorname{Tr}(K_1^2)\operatorname{Tr}(K_2^2)}}.
\end{equation}

This equation calculates a normalized Frobenius Inner-Product between the two matrices. This equation is just the cosine angle between the two kernels $K_1$ and $K_2$ if seen as vectors in the space of matrices. Thus, it is also commonly referred to as the ``geometric distance" (GD). As it is normalized, the alignment is bounded between 0 and 1, or $KA(K_1,K_2) \rightarrow [0,1]$.
This concept can be further specialized to calculate the alignment between the kernel, $K_1$ and the data labels it is classifying, $y$. The kernel-target alignment function~\cite{Wang2015} is defined as follows:

\begin{equation}
\label{eq:targetalignment}
    \operatorname{KTA}_{\boldsymbol{y}}(K,y)
      = \frac{\operatorname{Tr}(K \boldsymbol{y}\boldsymbol{y}^T)}{\sqrt{\operatorname{Tr}(K^2)\operatorname{Tr}((\boldsymbol{y}\boldsymbol{y}^T)^2)}}
      = \frac{\boldsymbol{y}^T K \boldsymbol{y}}{\sqrt{\operatorname{Tr}(K^2)} N}
\end{equation}

This function performs the kernel alignment of Equation \ref{eq:polarity} with an outer product of the labels, $K_2 = yy^T$. The higher the target alignment, the better the kernel fits the dataset vector in the feature space. If there is sufficient positive separation between the quantum and classical kernel target alignments, the kernel likely has a ``quantum advantage"~\cite{Huang2021}.

\subsection{Related Works in QML-based Anomaly Detection}
Existing QML literature supports the use of quantum anomaly detection models for other domains. Quantum CPS-specific anomaly work remains scarce, with only a few works mentioning CPS data in their proposed solutions. In recent work~\cite{jullian2022}, simple QSVM models were used on wind turbine systems. Furthermore, a gas power plant dataset was used to measure the effectiveness of a proposed autoencoder-reduced, parametrized QSVM feature map~\cite{Sakhnenko2022}.

\begin{table}[htbp]
\caption{Recent Works in Anomaly Detection with Quantum Machine Learning}
\centering
\label{table:comparison}
\resizebox{1.01\columnwidth}{!}{
    \setlength{\tabcolsep}{2pt}
    \renewcommand*{\arraystretch}{1.5}
    \begin{tabular}{cccccc}
    \toprule
        \makecell[c]{\bf Author and\\ \bf Year} & \makecell[c]{\bf QML\\ \bf Algorithm} & \bf Dataset & \makecell[c]{\bf \# of \\ \bf Qubits} & \bf Feature & \bf \makecell[c]{Metrics\\ (best)} \\
        \midrule
        
        \makecell[c]{Gouveia~et al.~\cite{network2}\\ 2020} & QSVM & \makecell[c]{NF-UNSW-\\NB15} & 16 & 16 & \makecell[c]{Acc: 64}\\
        \midrule
        
        \makecell[c]{Jullia~et al.~\cite{jullian2022}\\ 2022} & QSVM & WTS & 8 & 16 & \makecell[c]{Acc: 88.8\% \\ F1: 0.893}\\
        \midrule
        
        \makecell[c]{Tscharke~et al.~\cite{tscharke2023}\\ 2023} & QSVR & KDD & 5 & 5 & \makecell[c]{Acc: 82.0\% \\ F1: 0.78}\\
        \midrule
        
        \makecell[c]{Wang~et al.~\cite{MNIST}\\ 2023} & QHDNN & \makecell[c]{Fashion\\MNIST} & 16 & 16 & \makecell[c]{AUC: 88.24}\\
        \midrule
        
        \makecell[c]{Wang~et al.~\cite{MNIST}\\ 2023} & QHDNN & MNIST & 16 & 16 & \makecell[c]{AUC: 89.41}\\
        \midrule
        
        \makecell[c]{Kukliansky~et al.~\cite{network}\\ 2024} & QNN & \makecell[c]{NF-UNSW-\\NB15} & 16 & 16 & \makecell[c]{F1: 0.86}\\
        \midrule

        \makecell[c]{Cultice~et al.~\cite{Cultice2024}\\ 2024} & QOC-SVM & HAI & 8 & 16 & \makecell[c]{F1: 0.86 \\ ACC: 87\%}\\

        \bottomrule
    \end{tabular}
    }
\end{table}

An overview of recent research on anomaly detection with Quantum Machine Learning (QML) is given in Table~\ref{table:comparison} categorized by methods. Several research used QSVM-based approaches, including those by Gouveia~et al.~\cite{network2}, Correa-Jullia~et al.~\cite{jullian2022}, and Tscharke~et al.~\cite{tscharke2023}, with datasets such as NF-UNSW-NB15, WTS, and KDD. Wang~et al.~\cite{Wang2015} investigated hybrid models (QHDNN) on the Fashion MNIST and MNIST datasets, demonstrating competitive AUC values. QNN and QOC-SVM were employed in more recent research by Kukliansky~et al.~\cite{network} and Cultice~et al.~\cite{Cultice2024} respectively, and achieved good F1-scores on the NF-UNSW-NB15 and HAI datasets. The table illustrates how QML techniques are becoming more varied and efficient in anomaly detection tasks.

\section{Cyber-physical System Datasets} \label{sec:Datasets}
This work's example datasets all focus on industrial control systems related to water treatment, water distribution, and hydro-power turbine generation. A UMAP (Uniform Manifold Approximation and Projection) projection of each dataset is shown in Figure~\ref{fig:proj}. UMAP transforms high-dimensional data, which helps to visualize the datasets. 

The first dataset of this work is the Secure Water Treatment Testbed, or SWaT, by the iTrust, Centre for Research in Cyber Security, Singapore University of Technology and Design~\cite{Goh2017}. This 2015 dataset is described as a modern scaled-down, industry-compliant water treatment plant ~\cite{Qin2020}.
The dataset consists of collected network traffic and raw values of 25 sensors and 26 actuators. SWaT consists of 11 days of continuous operation of the emulation testbed, with 4 days of attacks. 41 unique combinations of attacks were launched. These attacks were thoroughly crafted through attack models designed by the iTrust research team. 

\begin{figure*}[htb]
\centering
\includegraphics[width=0.9\textwidth]{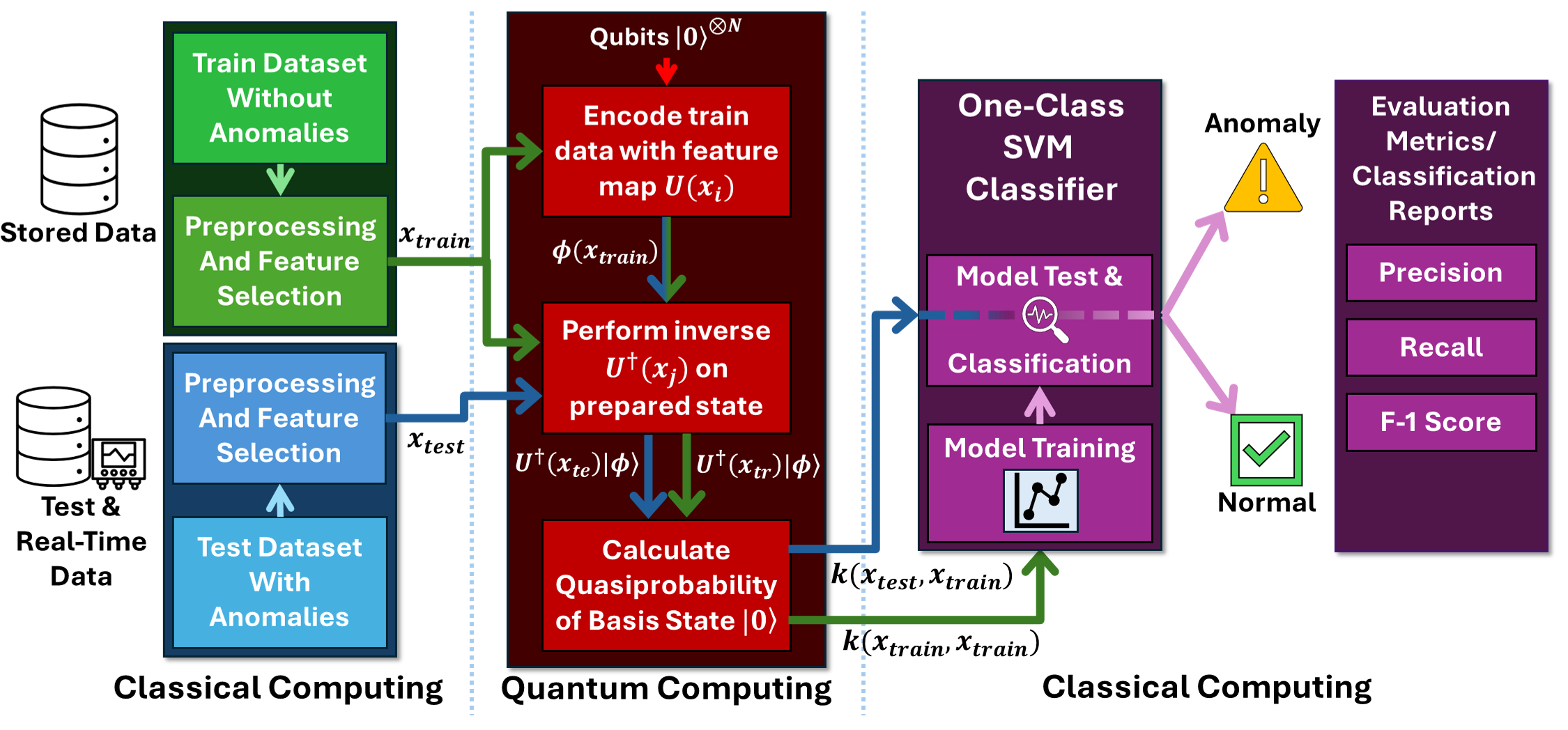}
\caption{Pipeline of the hybrid-quantum SVM anomaly detector. pre-processing occurs classically, is turned into a fidelity inner-product kernel using a quantum computer, and then utilized for SVM anomaly detection. Data kernels can then be passed into the trained SVM for detection of anomalies.}
\label{fig:overallFlow}
\end{figure*}

The second dataset we utilized in this work is the Hardware-in-Loop-based (HIL) Augmented Industrial Control System (ICS) Security Dataset or HAI 20.07~\cite{hai2020}. The HAI dataset is a realistic ICS testbed that physically emulates hydro-power and steam-turbine power generation.
In this dataset, there are 59 total features spread across all four SCADA processes.
Training data was captured from normal operation with no anomalies for 177 hours. In the test set, 38 attacks were conducted, including 14 different attack primitives and combinations of attacks.

The third and final dataset we used is the Water Distribution dataset, or WADI, by the iTrust, Centre for Research in Cyber Security, Singapore University of Technology and Design~\cite{Ahmed2017}. This dataset from 2017 recorded attacks on a testbed emulating a city water distribution network.
WADI contains 123 sensors and actuators between the three processes connected to the network. The dataset consists of 16 days of continuous operation, with two days had attack scenarios. 15 combinations of attack models were performed.

\section{Methodology} \label{sec:Methodology}
While there are many components of the Quantum-Hybrid Support Vector Machine (QSVM) that can be parametrized and interchanged, the overall algorithmic flow remains the same. It can be described as follows: Prepare the data and create train/test sets, calculate the quantum fidelity by performing the Hilbert-Schmidt inner product between the train data and current data vectors, and use the resulting similarity matrix to train and classify on the SVM's hyperplane. This process flowchart can be seen in Figure \ref{fig:overallFlow}.

The SVM's kernel takes advantage of quantum feature maps within the kernel function to identify hidden similarities or differences in data patterns that may be too computationally expensive for classical computing. To reduce the noise and resource consumption, the data pre-processing and SVM components are performed classically.

In previous work~\cite{Cultice2024}, this topic expanded upon only a single method of feature selection and one well-defined feature map. To get a clearer picture of the advantages, our current methodology parametrizes multiple aspects of the QSVM with known high-performing alternatives to identify the best.

\subsection{Data Pre-Processing and Feature Reduction}
Each dataset is prepared for the feature selection using some consecutive pre-processing techniques. Firstly, the dataset is scanned with a moving average of 60 overlapping sample windows. This helps reduce the noise and artifacts capturing the time series trends. Equation, $ma_t = \frac{1}{w} \times \sum_{i=t-w+1}^t x_i$ is used for the moving average where $w$ is the window length and $x_i$ is the sample at time $i$.

Next, the dataset is standardized with standard-scaler transformation or Z-score normalization with Equation~\eqref{std}. Here, $\mu$ is the sample mean and $\sigma$ is the standard deviation. This moves the mean of the distribution to zero and the standard deviation to 1, which helps to equalize the scales of different units. Furthermore, standardization tends to bring faster convergence for optimization techniques like gradient descent.
\begin{equation}
    x_{std} = \frac{x_o - \mu}{\sigma}
    \label{std}
\end{equation}

The pre-processing steps end with the feature reduction techniques. We experimented with three types of feature reduction techniques: a) Decision Tree, b) Principal Component Analysis (PCA), and c) Non-Negative Matrix Factorization(NMF). The number of features is chosen based on the number of qubits.

\begin{enumerate}[label=\alph*)]
    \item Decision Tree: Tree-based methods calculate the feature importance rank from the Gini impurity. Based on the relative decrease in the impurity it can sort the features from most important to less. 
    
    \item Principal Component Analysis: The PCA-based methods calculate the principal component from the dataset in eigenvalue representation. The eigenvalue defines the weight of that feature and can help rank it based on the relative weights.

    \item Non-Negative Matrix Factorization: NMF-based methods help reduce the number of features by decomposing a high-dimensional non-negative data matrix into two lower-dimensional non-negative matrices.
\end{enumerate}

The pre-processing techniques were applied separately to the train set and test set to avoid data leakage.

\subsection{Quantum Fidelity Kernel}
The kernel function is the most important component of a quantum-hybrid SVM, as it directly takes advantage of quantum computation. To create a quantum kernel, we use the quantum fidelity algorithm paired with a strong data-encoding Feature Map (FM) circuit. Proper parameter selection of the feature map is required for optimal class separation, including varying the repetitions, scale, and interactivity.

Each feature map encodes $N$ features in $n/2$ qubits to the Hilbert space using two degrees of freedom. Feature maps are also repeated 1-3 times to drive more interaction between features. Additionally, data is scaled between $[-\pi,\pi]$, with careful consideration into the encoded data's kernel bandwidth~\cite{Shaydulin2022}. A kernel inner-product matrix is created by fidelity testing each datapoint to one another, or $O(\lvert x_{input} \rvert \times \lvert x_{train} \rvert) $ calculations. Finally, the resulting kernels are post-processed by exponentiating the values, $k_p(x_i,x_j) = e^{k(x_i,x_j)}$.

In this work, we benchmark four separate feature maps from popular existing literature: Belis et al.~\cite{Belis2024}, Sakhnenko et al.~\cite{Sakhnenko2022}, the 2nd-Order Pauli-Z Feature Map~\cite{Havlicek2019}, and a simple U2-gate 2DoF feature map (for reference). These feature maps have been modified to take advantage of two degrees of freedom.

\subsubsection{Belis et al. Feature Map}
\begin{figure}[htbp]
\includegraphics[width=\linewidth]{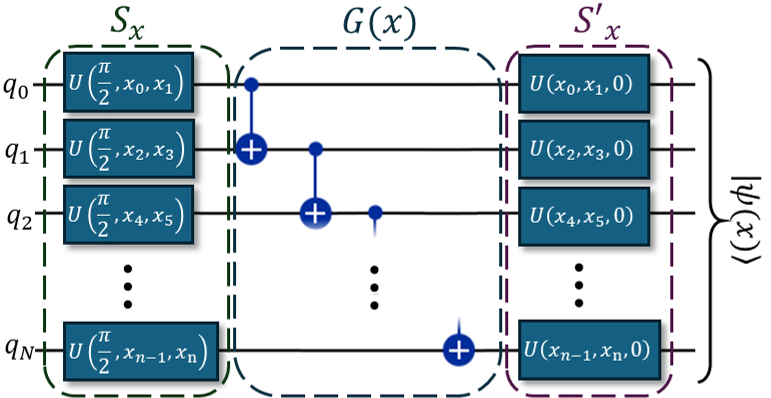}
\caption{2-DoF Belis et al.~\cite{Belis2024} Feature Map. $S_x$ and $S'_x$ angular encode $N$ features onto $N/2$ qubits, and $G(x)$ performs NN entanglement.}
\label{fig:belis_figure}
\end{figure}

This feature map proposed by Belis et al.~\cite{Belis2024} features encoding of two degrees of rotational freedom followed by nearest neighbor (NN) entanglement, $G(x)$, and a final rotation $S'_x$ on different axes. The gate $U(\pi/2,\phi,\lambda) \in SU(2)$ is a universal 1-qubit gate rotating about two axes. The circuit diagram is presented in Figure \ref{fig:belis_figure}. We repeat the gates of this feature map up to three times in the fidelity calculation.

\subsubsection{Simple ``2DoF" Rotation Feature Map}
As a ground state to determine how much the entanglements actually change the outcome, we have included a basic two-degree-of-freedom quantum feature map. This circuit consists of just universal 1-qubit $U(\pi/2,\phi,\lambda)$ gates which encode two features each onto $N/2$ qubits. There is no $G(x)$ entangling/interaction circuit, making this map extremely simple.

\subsubsection{Sakhnenko et al. Feature Map}
\begin{figure}[htbp]
\includegraphics[width=\linewidth]{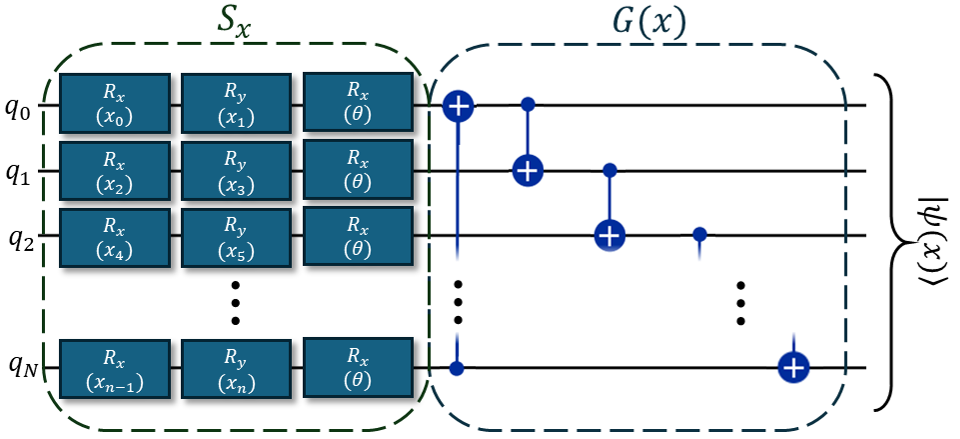}
\caption{2-DoF version of Circuit 10 from Sakhnenko et al.~\cite{Sakhnenko2022}}
\label{fig:sakh_figure}
\end{figure}

Feature maps based on parametrized circuit structures were proposed in Sakhnenko et al.~\cite{Sakhnenko2022} demonstrating strong results in gas turbine synthetic data. Particularly, their ``Circuit 10" performed very strongly, made up entirely of rotate x ($R_x$), rotate y ($R_y$), and CNOT gates. The angle parameter, $\theta$, was set to $\theta=\pi/2$ for all cases. We have created a 2-DoF variant inspired by the original Circuit 10 that is not repeated. This two rotational degree version of this circuit is provided in Figure \ref{fig:sakh_figure}.

\subsubsection{Second-Order Pauli-Z Evolution Feature Map}
\begin{figure}[htbp]
\includegraphics[width=\linewidth]{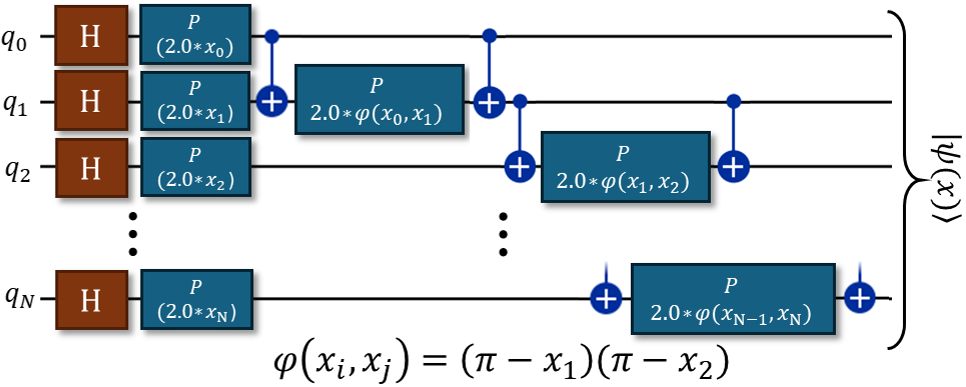}
\caption{1DoF, Second-Order Pauli-Z Evolution (ZZ Feature Map)~\cite{Havlicek2019}.}
\label{fig:zzfm_figure}
\end{figure}

The ZZ Feature Map (ZZFM) is the most popular method of data encoding which uses 2nd-order Pauli-Z evolution with linear entanglement~\cite{Havlicek2019}. This feature map presented in Figure \ref{fig:zzfm_figure} utilizes a classical non-linear function, $\varphi(x,y) = (\pi - x)(\pi - y)$, as parameters in Pauli gates. Unlike the others, ZZFM utilizes 1DoF, or $N$ qubits for $N$ features. It also does not contain separated encoding and entanglement operations, choosing to interweave them instead. This was implemented using Qiskit's \textit{ZZFeatureMap}~\cite{qiskit2024} construct.

\subsection{One-Class Support Vector Machine Classification}
One-Class Support Vector Machines (OCSVM) are a robust and widely used method for anomaly detection. They are a variant of Support Vector Machines (SVM) originally proposed by work by Cortes and Vapnik~\cite{svm_1} in 1995 where they discussed the optimal hyperplane in high dimensional feature space. This is particularly useful where the training data predominantly consist of baseline or normal observations without many anomaly data points.

The kernel trick is a powerful method to handle non-linearly separable data by implicitly transforming it into a higher-dimensional space where it becomes more linearly separable. Instead of explicitly computing the transformation, the kernel trick uses a kernel function to compute the transformed vectors directly. For example, common classical kernel functions include the linear, polynomial, and radial basis function (RBF) kernels.

The OCSVM learns the decision function $f(x)$ in such a way that most of the normal class data points lie within the decision boundary. It uses the ``kernel trick" to elevate the data points and map them in high-dimensional feature space. With the increase in dimension, it becomes easier for the classifier to distinguish between classes. The optimization function is:
\begin{equation}
    \underset{w,\rho,\zeta}{min}\left[\frac{1}{2}\|w\|^2 + \frac{1}{b\times n} \times \sum_{i=1}^N \zeta_i - \rho\right]
\end{equation}

with the constraint:
\begin{equation}
    \left (w\times \phi(x) \right ) \leq \rho - \zeta_{i}, where\ \zeta_{i}>0,\ and\ i = 1, 2, ..., n
\end{equation}
The final decision function is:
\begin{equation}
    f(x) = \left (w\times \phi(x) \right ) - \rho
\end{equation}
Here, $\phi(x)$ is the feature mapped data, $w$ is the normal vector to the hyperplane, $\phi$ defines the offset for the hyperplane, $\zeta$ ensures the soft decision boundary by allowing a certain number of anomalous points, and $\nu$ is the parameter that controls the outliers by setting an upper bound.\\
The training process solves the following optimization problem by taking the kernel $k(x_i,x_j)$ into account, where $k(x_i,x_j) = \phi(x_i)\cdot\phi(x_j)$.
\begin{equation}
    \underset{\alpha}{max}\left[-\frac{1}{2}\sum_{i,j=1}^N\alpha_i\times\alpha_j\times k(x_i,x_j)  \right]
\end{equation}
Given, $0\leq\alpha\leq\frac{1}{\nu\times n}$ and also, $\sum_{i=1}^n\alpha_i = 1$, where, $\nu$ is the Lagrange multiplier. In this work, the kernel function $k(x_i,x_j)$ is computed by the feature map fidelity algorithm defined earlier.

% SWAT Graphs
\begin{figure*}[htbp]
    \begin{subfigure}[h]{0.5\linewidth}
    \includegraphics[width=\linewidth]{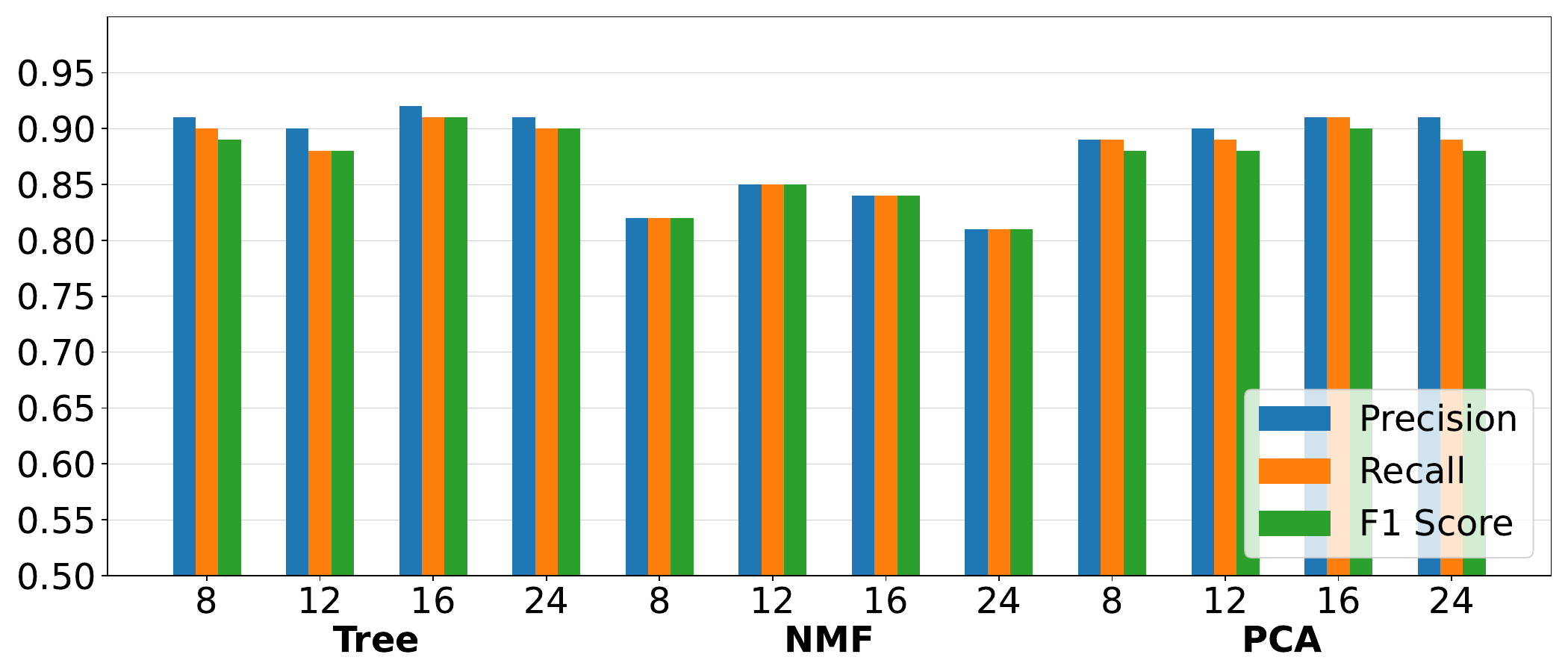}
    % trim={2.5cm 0 3cm 1.8cm},clip,
    \caption{Belis Feature Map Ideal Performance}
    \label{fig:quantum_swat_belis}
    \end{subfigure}
\hfill
    \begin{subfigure}[h]{0.5\linewidth}
    \includegraphics[width=\linewidth]{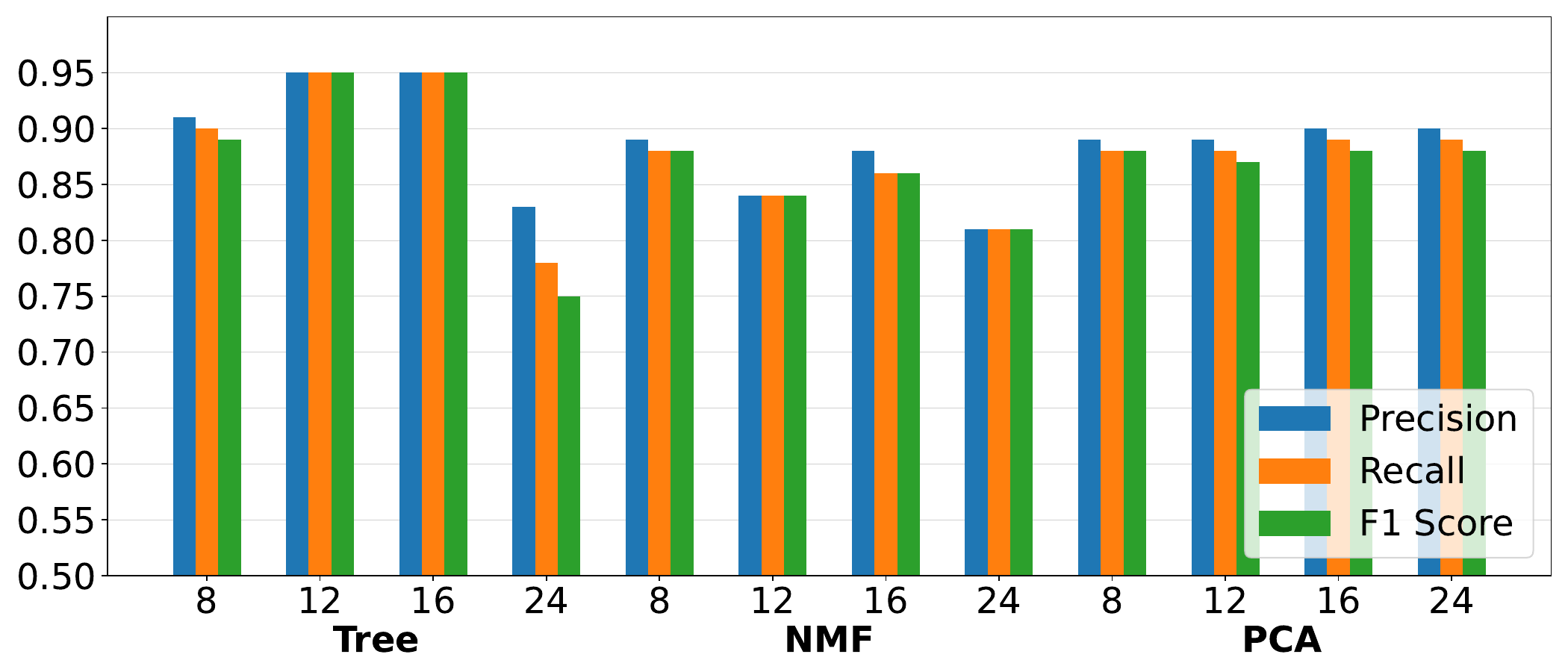}
    % trim={2.5cm 0 3cm 1.8cm},clip,
    \caption{Basic 2DoF Feature Map Performance}
    \label{fig:quantum_swat_2dof}
    \end{subfigure}%
    \\
    \begin{subfigure}[h]{0.5\linewidth}
    \includegraphics[width=\linewidth]{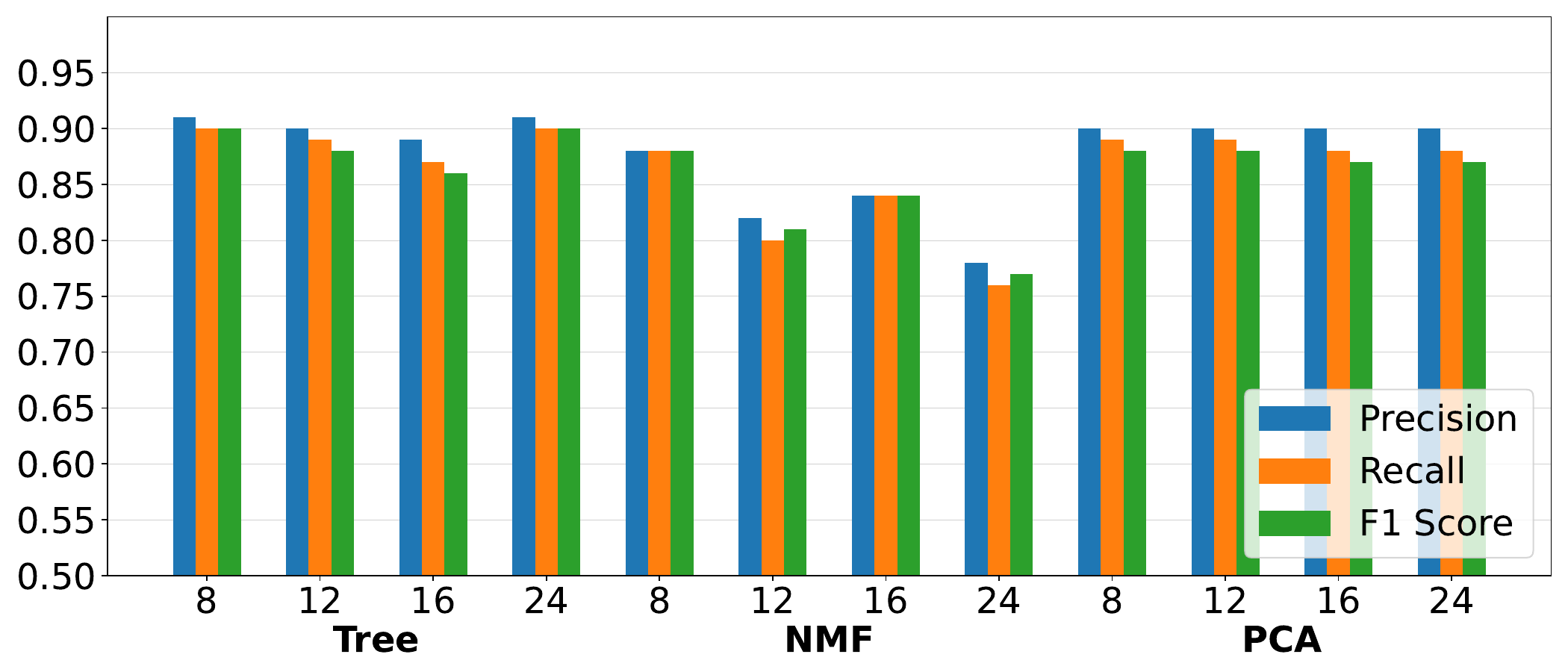}
    % trim={2.5cm 0 3cm 1.4cm},clip,
    \caption{Sakhnenko Feature Map Performance}
    \label{fig:quantum_swat_sakh}
    \end{subfigure}
\hfill
    \begin{subfigure}[h]{0.5\linewidth}
    \includegraphics[width=\linewidth]{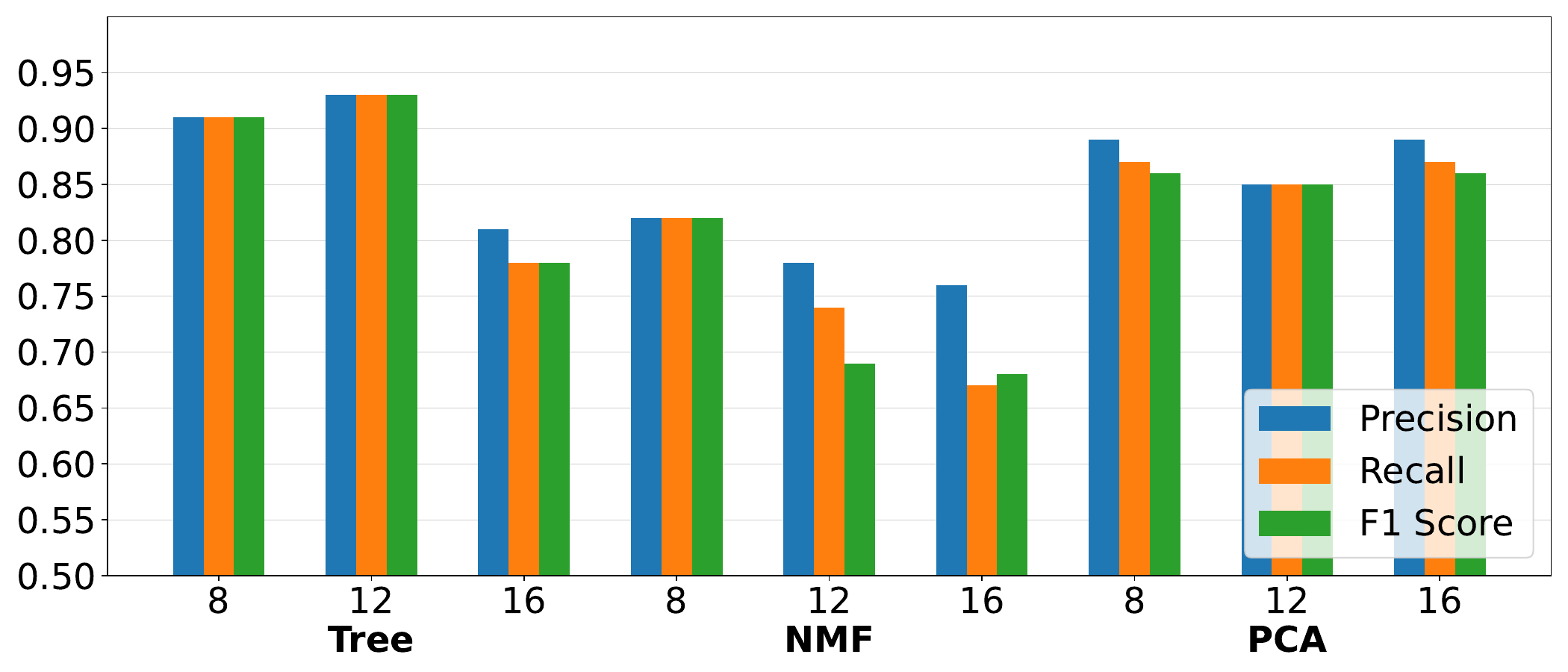}
    % trim={2.5cm 0 3cm 1.4cm},clip,
    \caption{ZZ Feature Map Performance}
    \label{fig:quantum_swat_zzfm}
    \end{subfigure}%
\caption{Results for Quantum SVM Anomaly Classification on SWaT for all feature maps.}
\label{fig:quantum_swat}
\end{figure*}

\section{Experimentation Results} \label{sec:Results}
To determine the effectiveness of quantum kernels in the ICS ecosystem, thoughtful experimentation was conducted to quantify the benefits of the highly non-linear behaviors of quantum computing. This experimentation investigates different pre-processing methods, feature maps, and features. The SVM parameters less relevant to concluding our work, such as the nu and tolerance, were swept and best results selected to ensure all classical and quantum results were selected fairly and optimally. 

The Belis, Basic 2DoF, and Sakhnenko feature maps were parametrized between three pre-processing methods (Decision Tree, NMF, and PCA), and four feature counts ($N=8,12,16,24$). Due to the exponential computing complexity, the ZZ Feature Map was only used for $N=8,12,16$.

\subsection{Dataset Results: SWaT}
The results, as shown in Figure \ref{fig:quantum_swat}, start off strong with very good classification performance from quantum kernels on the SWaT dataset. The three 2DoF feature maps all reach F1 metrics of over 0.9 with multiple confligurations. While the feature maps had some degraded performance on the NMF feature reduction method, the performance remains acceptable for fairly reliable classification tasks.

The basic 2DoF feature map had the best performance of 0.95 F1 score with the Decision Tree feature selection of 16 features. Thus, we can theorize that the entanglement structures of the other feature maps did not provide any meaningful improvement over simply encoding with universal rotation gates. In fact, the ZZ Feature Map circuit struggled with consistency, varying in performance over the different pre-processing methods.

\subsection{Dataset Results: HAI}
The HAI dataset was arguably the most difficult to classify due to its very complex, non-linear relationships between features and noisy data. Similarly to previous work~\cite{Cultice2024}, the selection of features using a Gini impurity decision tree was the best method of pre-processing for quantum kernels. This was likely due to many inconsequential features that have very low correlation to each other or trends in the data. 

All feature maps averaged above 0.82 in all three classification scores with the decision tree, which is well above the average of the other pre-processing methods. The best results showed in Belis kernel, 16 tree-reduced features with 0.88/0.87/0.86 in precision, recall, and F1. In the Belis, 2DoF, and Sakhnenko kernels, PCA performed very poorly at an average of 0.63 in the classification metrics. However, the ZZ Feature Map's highly non-linear design actually could find good decision boundaries in the PCA-reduced HAI data. This resulted in performance similar to the decision tree at 0.8 precision, 0.8 recall, and 0.79 F1. This further suggests that the features pruned by the tree add unnecessary relational feature complexity with no benefit in classification. The NMF results were also very low, almost averaging close to a purely random chance of 50-50. Ultimately, the HAI dataset has shown that the method of reducing features and condensing data plays a huge role in classifiability, especially in quantum kernel bandwidth.

\begin{figure*}[htbp]
    \begin{subfigure}[h]{0.5\linewidth}
    \includegraphics[width=\linewidth]{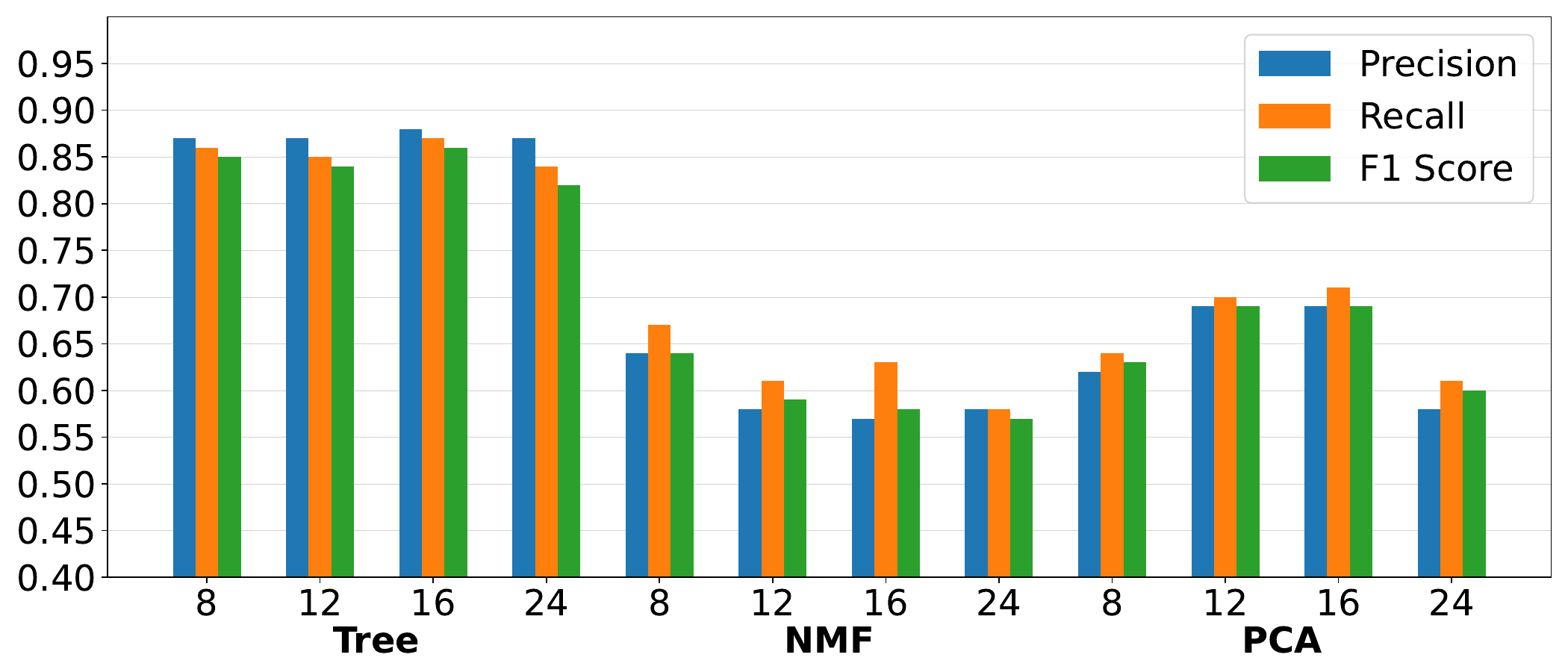}
    % trim={2.5cm 0 3cm 1.8cm},clip
    \caption{Belis Feature Map Ideal Performance} 
    \label{fig:quantum_hai_belis}
    \end{subfigure}
\hfill
    \begin{subfigure}[h]{0.5\linewidth}
    \includegraphics[width=\linewidth]{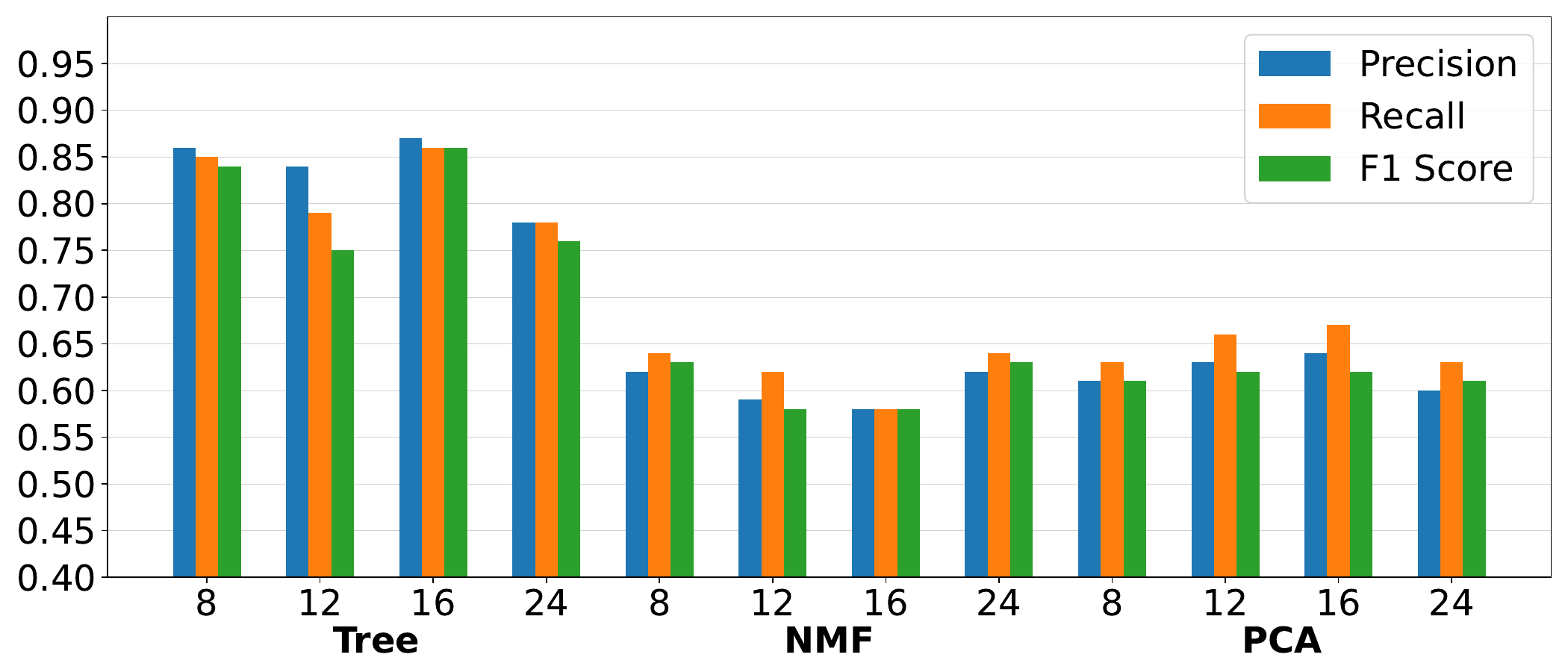}
    %trim={2.5cm 0 3cm 1.8cm},clip,
    \caption{Basic 2DoF Feature Map Performance}
    \label{fig:quantum_hai_2dof}
    \end{subfigure}%
    \\
    \begin{subfigure}[h]{0.5\linewidth}
    \includegraphics[width=\linewidth]{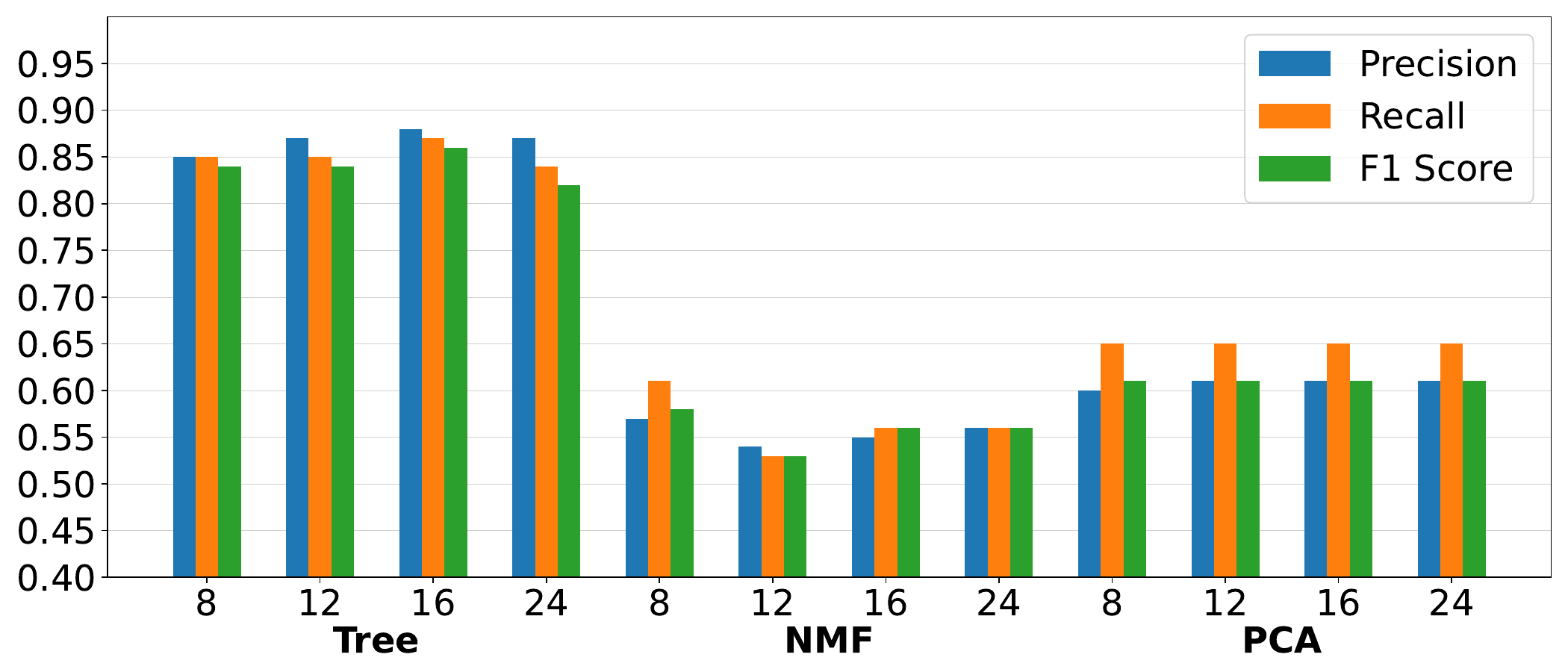}
    % trim={2.5cm 0 3cm 1.4cm},clip,
    \caption{Sakhnenko Feature Map Performance}
    \label{fig:quantum_hai_sakh}
    \end{subfigure}
\hfill
    \begin{subfigure}[h]{0.5\linewidth}
    \includegraphics[width=\linewidth]{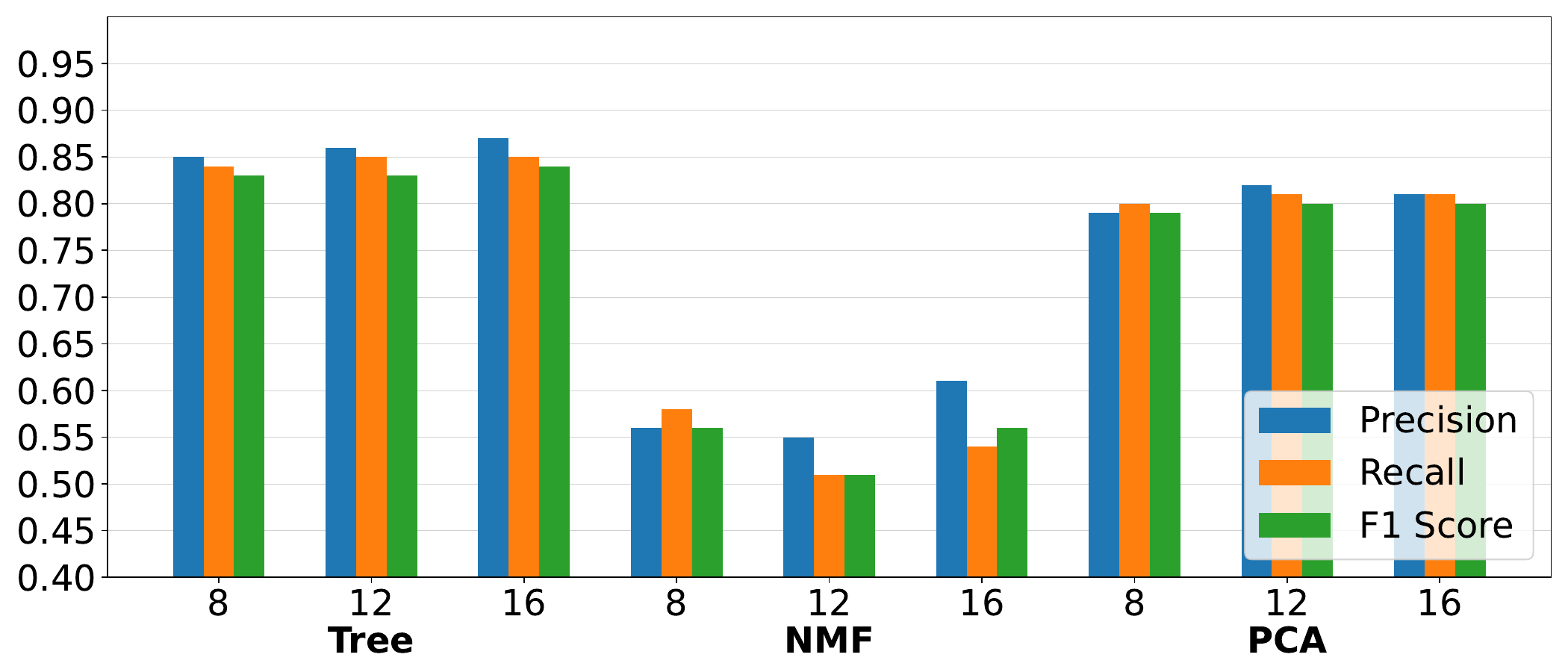}
    % trim={2.5cm 0 3cm 1.4cm},clip,
    \caption{ZZ Feature Map Performance}
    \label{fig:quantum_hai_zzfm}
    \end{subfigure}%
\caption{Results for Quantum SVM Anomaly Classification on HAI for all feature maps.}
\label{fig:quantum_hai}
\end{figure*}

\subsection{Dataset Results: WADI}
The WADI dataset was seemingly average in classification complexity and difficulty based on the results. Out of the three pre-processing methods, the Decision Tree selection method performed the best, likely due to many confounding/noisy features of the least importance being pruned by the selection algorithm. Similarly to before, the NMF pre-processing method was devastating for both quantum and classical methods, likely due to the dataset's particular distribution.

Most quantum feature maps thrived with the data, performing consistently above 0.7 Precision/Recall/F1 Score with all configurations. The maximum performing model was the simple 2DoF model using Decision Tree selection of $N=16$ features, reporting a Precision of 0.9, Recall of 0.89, and F1 of 0.89. It is highly likely that qubit entanglement did not greatly impact the kernel or SVM trained hyperplane, as the simple $U2$-gate feature map performed just as well as the entangled circuits. Furthermore, the ZZ Feature Map underperformed on this task, further reinforcing the idea that simple quantum models may have an advantage with this specific data.

\begin{figure*}[htbp]
    \begin{subfigure}[h]{0.5\linewidth}
    \includegraphics[width=\linewidth]{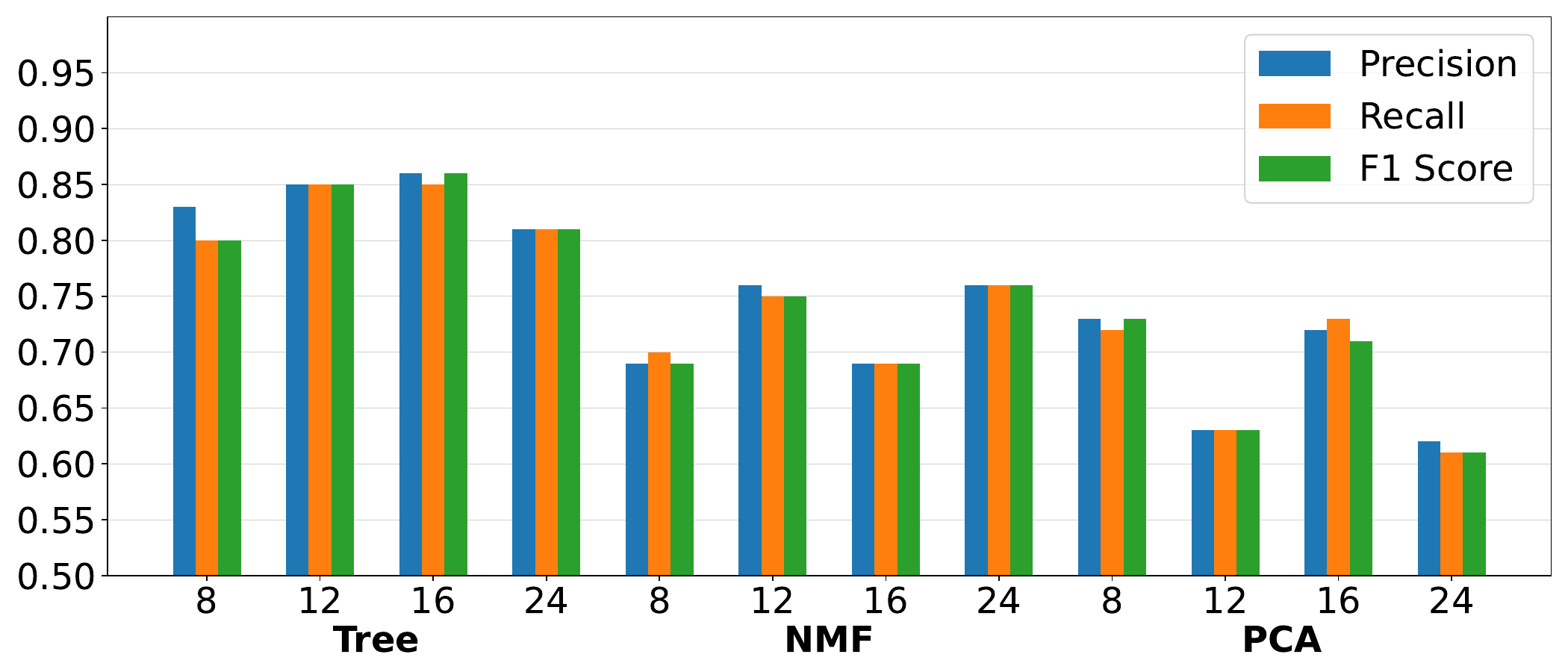}
    % trim={2.5cm 0 3cm 1.8cm},clip,
    \caption{Belis Feature Map Ideal Performance}
    \label{fig:quantum_wadi_belis}
    \end{subfigure}
\hfill
    \begin{subfigure}[h]{0.5\linewidth}
    \includegraphics[width=\linewidth]{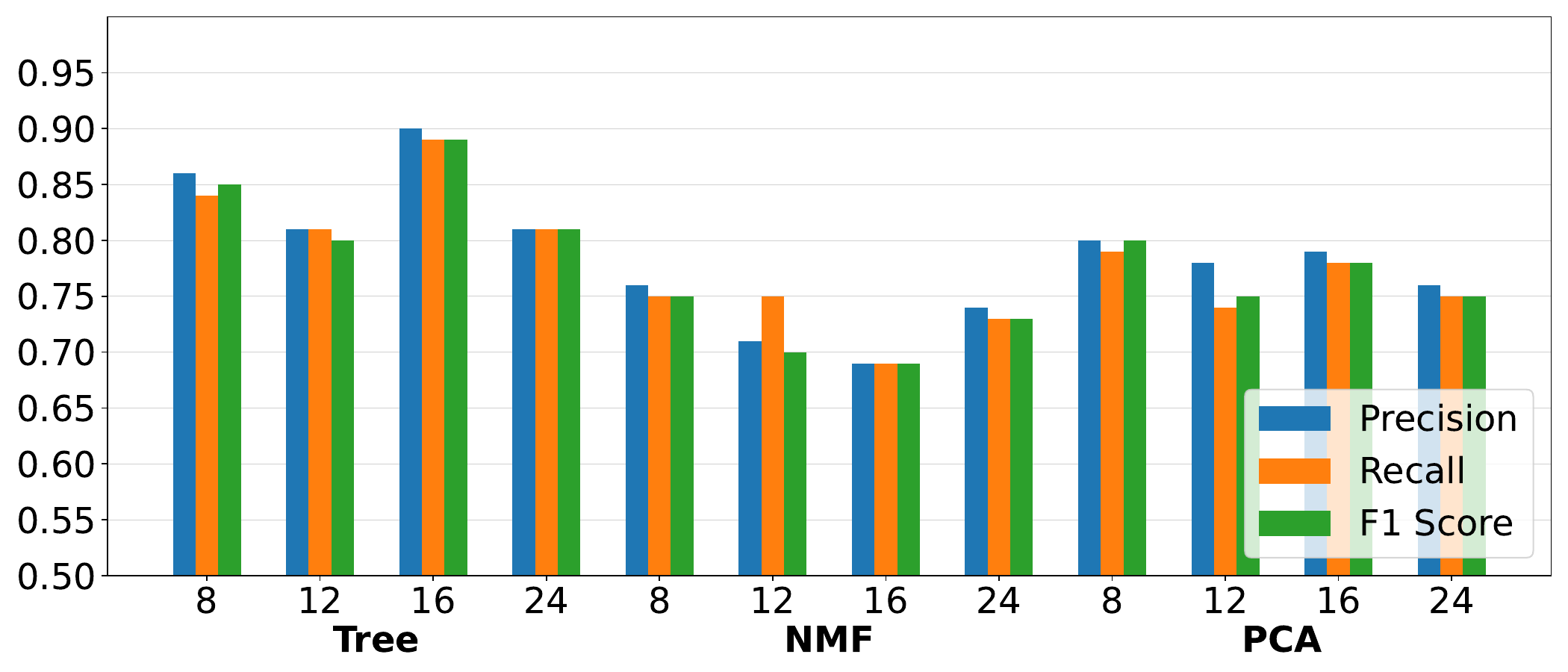}
    % trim={2.5cm 0 3cm 1.8cm},clip,
    \caption{Basic 2DoF Feature Map Performance}
    \label{fig:quantum_wadi_2dof}
    \end{subfigure}%
    \\
    \begin{subfigure}[h]{0.5\linewidth}
    \includegraphics[width=\linewidth]{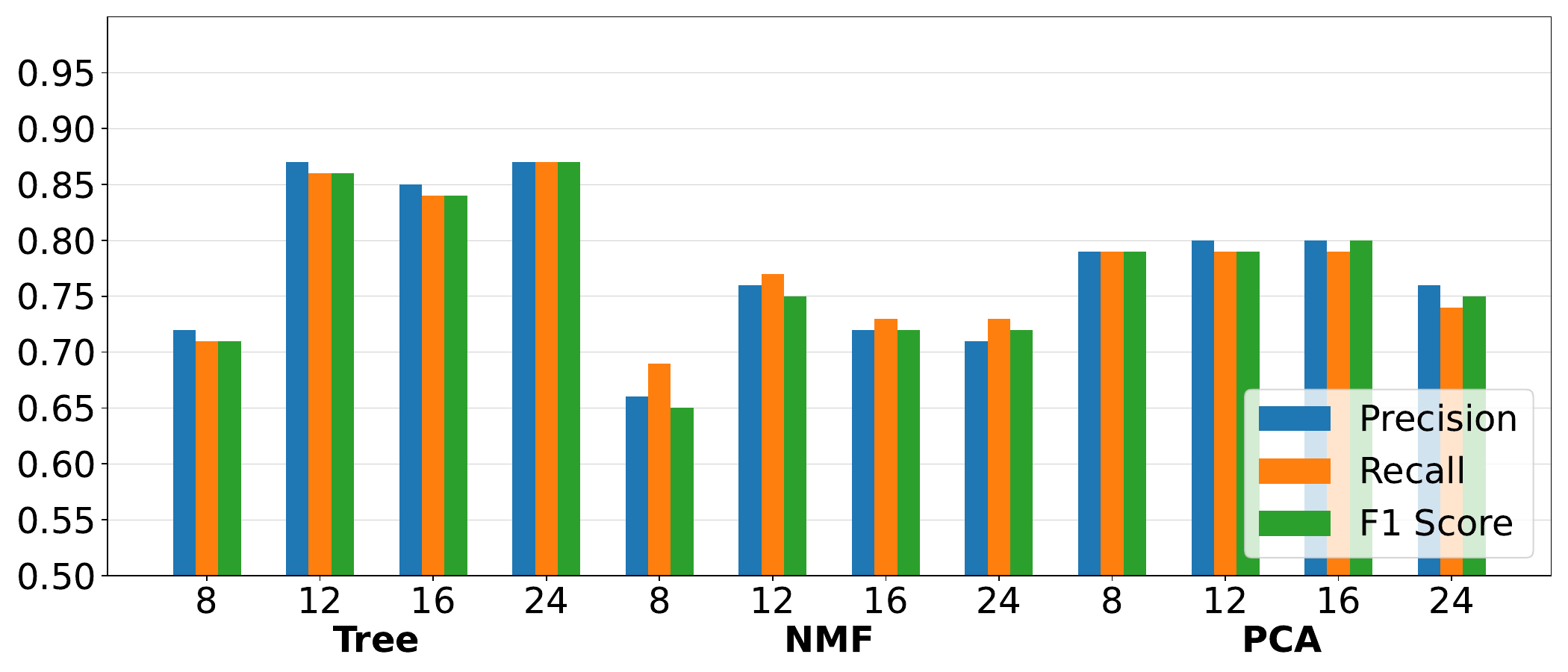}
    % trim={2.5cm 0 3cm 1.4cm},clip,
    \caption{Sakhnenko Feature Map Performance}
    \label{fig:quantum_wadi_sakh}
    \end{subfigure}
\hfill
    \begin{subfigure}[h]{0.5\linewidth}
    \includegraphics[width=\linewidth]{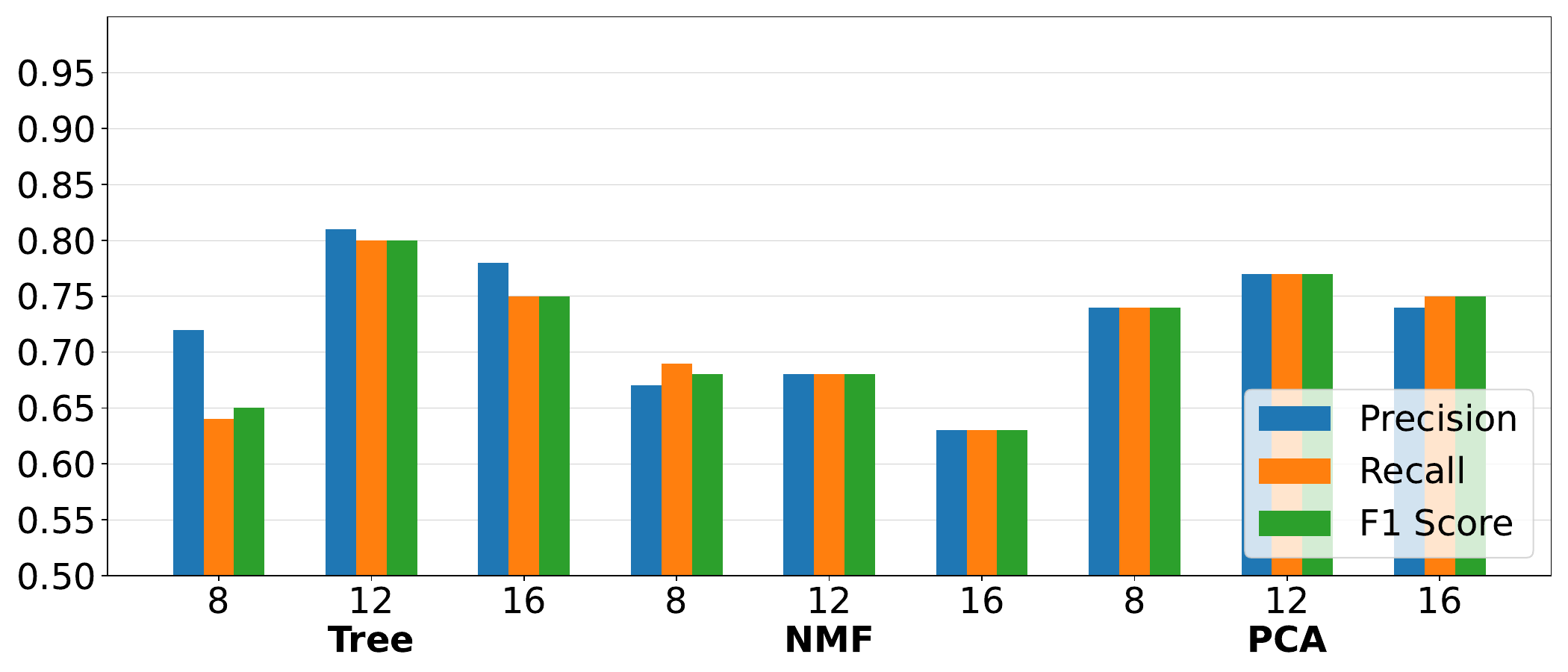}
    % trim={2.5cm 0 3cm 1.4cm},clip,
    \caption{ZZ Feature Map Performance}
    \label{fig:quantum_wadi_zzfm}
    \end{subfigure}%
\caption{Results for Quantum SVM Anomaly Classification on WADI for all feature maps.}
\label{fig:quantum_wadi}
\end{figure*}

\subsection{Classical Results}

\begin{figure}[htbp]
\includegraphics[width=\linewidth]{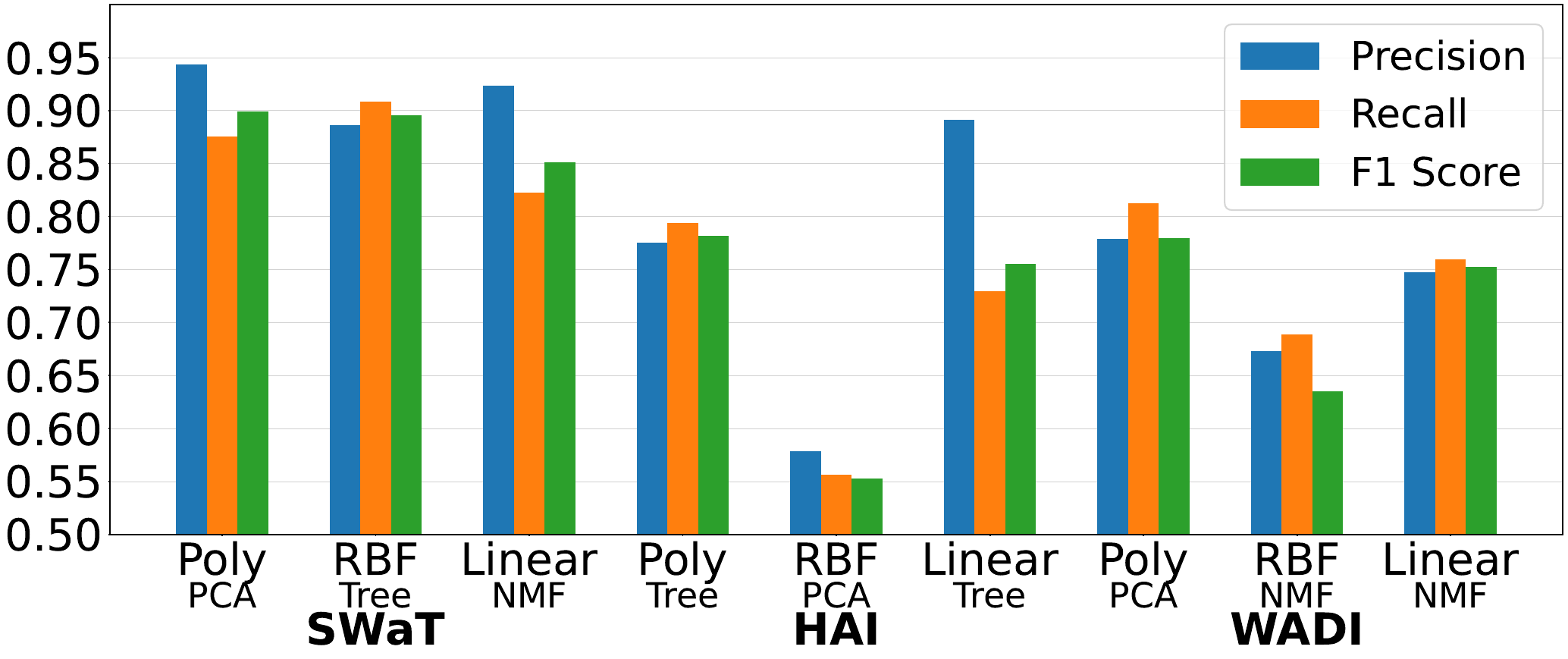}
\caption{Best Results for Classical Kernels' SVM Classification on all datasets.}
\label{fig:classical_all}
\end{figure}

To identify any improvement to classification, classical kernel function results are presented for Polynomial, RBF, and Linear kernels in Figure \ref{fig:classical_all}. Overall, classical provided consistently lower results with more variation than quantum for all three datasets. For the SWaT dataset, we see a similar performance between the two, with an average of 6\% difference in metrics. The best classical result for SWaT was the RBF kernel using a decision tree, which achieved a recall of 0.91 and F1 of 0.89. However, classical averaged 0.88 F1 score across all kernel and preprocessing configurations, leading us to conclude that the SWaT dataset is overall just easier to classify compared to the other CPS datasets.

% Classical Graphs

For HAI, classical methods struggled on NMF and PCA data, quite similar to the quantum kernels. However, the models based on polynomial/linear dot products were able to produce acceptable results using the decision tree method, with the best result of 0.77/0.79/0.78 in precision, recall, and F1. Supported by previous work~\cite{Tushkanova2023}, this leads us to believe that HAI has many confounding features as the decision tree would remove low-quality information. However, this is still over 10\% lower than the best quantum kernel. Quantum kernels still significantly outmatched the classical, averaging approximately 4.6\% higher in precision, 9.8\% higher in recall, and 11.8\% higher in F1 score. Thus, we can conclude that using a quantum kernel has made significant improvements over classical in HAI.

 The WADI dataset resulted also in favor of the quantum kernel methods. The best quantum feature map configurations performed 13\% better over all classical functions in the classification performance of WADI. For this dataset, the NMF feature reduction method seemed to be optimal for classical kernels. On average, classical kernels presented a modest F1 score of 0.722. However, when factoring in noise, quantum outperformed classical by only 4.2\% on average, lowering the experimental certainty of a quantum advantage.
 
 Overall, quantum models performed an average of 13.3\% better than their classical counterparts across each preprocessed dataset and kernel. These results help us conclude that quantum kernels can perform consistently above classical methods in these particular CPS datasets. Common trends between classical and quantum suggest that certain datasets and kernel types favor specific pre-processing methods, such as the importance decision tree.

% IBM Quantum Computer Table
\begin{table}
\centering
\caption{Median errors and decoherence times in IBM Quantum computers captured at the time of noise model simulation.}
\label{tbl:ibm_noisy}
\setlength\tabcolsep{0.3em} {
\begin{tabular}{ccccccccc}
\toprule
  Instance & \# of & Best 2Q & SX & CZ/ECR & Readout & T1 & T2 \\
  Name & Qubits & EPLG & Gate & Gate & Asgmt. & (us) & (us) \\
\midrule
Torino & 133 & 1.08e-3 & 3.149e-4 & 2.936e-3 & 2.6e-2 & 177.93  & 137.79 \\
Sherbrooke & 127 & 2.39e-3 & 2.582e-4 & 7.726e-3 & 1.39e-2 & 262.43 & 164.33 \\
Kyiv & 127 & 4.33e-3 & 2.779e-4 & 1.208e-2 & 8.2e-3 & 265.84 & 101.59 \\
\bottomrule
\end{tabular}
}
\end{table}

\section{Quantum Error and Advantage Metrics} \label{sec:Results2}

% Quantum Average Error and Percent Decrease
\begin{table*}
\centering
\caption{Average error and percent decrease of classification metrics when using noisy kernel simulations.}
\label{tbl:ibm_noise_decrease}
\resizebox{2\columnwidth}{!}{
\renewcommand{\arraystretch}{1.4}
\setlength\tabcolsep{10pt}
\begin{tabular}{cc|ccc|ccc|ccc}
\hline
\multirow{3}{*}{\rotatebox{90}{Dataset}} & & \multicolumn{9}{c}{$\% \downarrow$ in Performance Metrics} \\
\cline{3-11}
 & Kernel & \multicolumn{3}{c|}{8 Features} & \multicolumn{3}{c|}{12 Features} & \multicolumn{3}{c}{16 Features} \\
\cline{3-11}
  & & Prec & Rec & F1 & Prec & Rec & F1 & Prec & Rec & F1\\
\hline
\multirow{2}{*}{\rotatebox{90}{SWaT}} & Belis~\cite{Belis2024} & 1.12\% & 2.20\% & 1.11\% & 0.00\% & 0.00\% & 0.00\% & 3.30\% & 4.35\% & 5.49\%  \\
& 2DOF & 0.00\% & 1.09\% & 0.00\% & 2.27\% & 1.11\% & 2.27\% & 1.10\% & 1.09\% & 1.10\% \\
\hline
\multirow{2}{*}{\rotatebox{90}{HAI}} & Belis~\cite{Belis2024} & 2.35\% & 2.30\% & 2.33\% & 0.00\% & 0.00\% & 0.00\% & 1.16\% & 1.14\% & 1.15\% \\
& 2DOF & 0.00\% & 0.00\% & 0.00\% & 0.00\% & 1.15\% & 0.00\% & 0.00\% & 0.00\% & 0.00\% \\
\hline
\multirow{2}{*}{\rotatebox{90}{WADI}} & Belis~\cite{Belis2024} & 2.50\% & 3.61\% & 2.50\% & 8.23\% & 9.41\% & 9.41\% & 15.12\% & 12.88\% & 14.11\% \\
& 2DOF & 1.25\% & 9.64\% & 11.25\% & 1.18\% & 4.71\% & 3.53\% & 3.49\% & 4.65\% & 5.88\% \\
\hline
\end{tabular}
}
\end{table*}

\subsection{Noise Effects on Classification Tasks}
Quantum noise was estimated from real IBMQ hardware in the quantum fidelity circuits. This includes three popular, modern machines: IBM Kyiv, IBM Sherbrooke, and IBM Torino. To simulate noise, we used Qiskit AER~\cite{qiskit2024} density matrix simulations with estimated noise snapshots from the time of experimentation, which are presented in Table \ref{tbl:ibm_noisy}. Due to the computational complexity of simulating noisy density matrices, noisy results were computed only for feature count $N = 8,12,16$ with the two best performing feature maps: Belis et al. \cite{Belis2024} and Simple 2DoF. The results, represented as percent decrease in classification metric between statevector and noisy simulations, are presented in Table \ref{tbl:ibm_noise_decrease}.

Noise did not strongly impact the classification of SWaT. Both kernels retained consistent results for 8 and 12 features, suffering at most 2.27\% percent change reduction in classification F-1. However, the Belis kernel experienced up to 5.49\% reduction in F1 score when using 16 features, likely due to the noisy disturbances pushing a significant cluster of datapoints beyond the hyperplane boundary. However, as this is only percent change, we can determine that the kernels were resilient to disturbances to noise and still retain an advantage of their classical counterparts.

For the HAI dataset, the effects of noise were even less impactful on our experimental conclusions. The simple benchmark 2DoF kernel was extremely resilient to change, with a 1.15\% decrease in results on 12-feature's recall. Similarly, the Belis kernel barely fluctuated, with less than 2.35\% reduction in classification quality across all feature counts. This lack of classification degradation leads to the conclusion that the HAI dataset clusters are fairly robust and changes in the kernel don't meaningfully change the resulting hyperplane.

The effect of noise was somewhat drastic for the WADI dataset. For the Belis Feature Map, the performance drops anywhere between 1\% and 15\% with error. For the basic 2DoF feature map, the negative effect of noise is similar, with up to 11.25\% decrease in F1 for 8 features. However, the overall average error of 2DoF on WADI is 3.57\% less than Belis, suggesting again that decoherence of longer circuits, especially repeated ones, definitely play a role in kernel variation. This coincides with WADI having the highest dissimilarity error in Section \ref{sec:noiseEffect} out of all three datasets.

\subsection{Noise Effects in Kernel Error} \label{sec:noiseEffect}
To understand the effects of noise on the quantum kernel, we define kernel error using the kernel geometric difference function described in Section \ref{sec:Background}. This allows us to measure the geometric similarities between the two kernels. However, as the goal is to capture the difference between the statevector (ideal) and noisy matrices, we represent our results as the complementary of the normalized alignment. Thus, we define $D_{Error} = 1-KA(K_{ideal},K_{noisy})$, where $D_{Error}$ is the dissimilarity between the matrices. This means a lower value ($D_{Error} \rightarrow 0$) is better, as the noisy kernel would have less deviation from the ideal model.

% Noisy Kernel KTA Error
\begin{table}
\centering
\caption{Reported error in noisy simulation kernels of $N$ features using dissimilarity metric: $1-KA(K_{ideal},K_{noisy})$.}
\label{tbl:noise_gd}
 \renewcommand{\arraystretch}{1.4}
\setlength\tabcolsep{0.3em} {
\begin{tabular}{cc|cc|cc|cc}
\hline
  \multirow{3}{*}{\rotatebox{90}{Dataset}}& \multirow{3}{*}{N} & \multicolumn{2}{c|}{\multirow{2}{*}{Torino}} & \multicolumn{2}{c|}{\multirow{2}{*}{Sherbrooke}} & \multicolumn{2}{c}{\multirow{2}{*}{Kyiv}} \\
  &  &  &  &  &  &  &  \\
  \cline{3-8}
  &  & Belis~\cite{Belis2024} & 2DOF & Belis~\cite{Belis2024} & 2DOF & Belis~\cite{Belis2024} & 2DOF \\
\hline
\multirow{3}{*}{\rotatebox{90}{SWaT}} & 8 & 3.67e-03 & 3.71e-03 & 1.46e-03 & 1.07e-03 & 9.73e-04 & 7.04e-04 \\
& 12 & 6.98e-03 & 1.06e-03 & 3.79e-03 & 6.34e-04 & 2.36e-03 & 3.41e-04 \\
& 16 & 8.60e-03 & 1.09e-03 & 6.91e-03 & 1.02e-03 & 3.07e-03 & 3.66e-04 \\
\hline
\multirow{3}{*}{\rotatebox{90}{HAI}} & 8 & 1.02e-03 & 8.68e-04 & 5.59e-06 & 1.82e-06 & 5.92e-04 & 8.98e-05 \\
& 12 & 6.50e-03 & 1.12e-03 & 7.06e-03 & 5.62e-04 & 3.61e-03 & 2.25e-04 \\
& 16 & 1.25e-03 & 1.07e-03 & 1.07e-05 & 2.11e-06 & 6.32e-06 & 1.30e-06 \\
\hline
\multirow{3}{*}{\rotatebox{90}{WADI}} & 8 & 4.71e-03 & 5.32e-03 & 1.28e-03 & 1.07e-03 & 9.84e-04 & 8.43e-04 \\
& 12 & 5.62e-03 & 6.44e-03 & 5.99e-03 & 3.00e-03 & 7.77e-03 & 4.00e-04 \\
& 16 & 3.59e-03 & 9.80e-03 & 3.26e-03 & 4.67e-03 & 4.12e-03 & 2.61e-04 \\
\hline
\end{tabular}
}
\end{table}

Based on the results presented in Table \ref{tbl:noise_gd}, we can see that the geometric dissimilarity is relatively low, allowing our noisy results to closely resemble the ideal results. This resilience to quantum noise can be both in part of mitigated error from IBM circuit compilation and our low-depth circuits. Despite this, though, even the smallest of kernel perturbations can cause immense classification troubles if the hyperplane is instable, as shown with the noisy results of WADI.

Interestingly, the effect of the Readout Assignment Error from Table \ref{tbl:ibm_noisy} seems to have correlation to the impact on the results. For example, IBM Torino's Readout error at the time of experimentation was $2.6\times10^{-2}$, which was 2.78 and 1.87 times greater than IBM Kyiv and IBM Sherbrooke, respectively. We can see a very similar ratio in the IBM machine's dissimilarity error of Table \ref{tbl:noise_gd}, the differences increasing rapidly for larger values of $N$ features. This effect is also reflected in the error bars of the classification performance.

Furthermore, we can see the expected effect of gate and relaxation errors on longer circuits as the dissimilarity of the Belis feature map is often more deviated/errored than the basic 2DoF feature map. Longer circuits mean more chance of gate and decoherence issues, which is present in the dissimilarity. This most assuredly should be explored further in future work.

\subsection{Identifying a Quantum Advantage}
``Quantum advantage" is a term used in literature to describe if there is substantial evidence to claim the benefits of using Quantum over classical ML. To find this, we must determine how much benefit performing high-dimensional feature mapping in quantum provides over well-known classical methods. The best way to identify this quantum advantage is to quantify the ``target-kernel alignment" defined in Section \ref{sec:Background} between the validation set and kernel matrices~\cite{Huang2021}. We used the PennyLane~\cite{Pennylane} implementation of the target alignment to calculate this similarity metric and presented it in Table \ref{tbl:quant_advantage}.

% Kernel Target Alignment Table
\begin{table}
\centering
\caption{Kernel Target Alignment of quantum and classical kernels to each pre-processed dataset. }
\label{tbl:quant_advantage}
\renewcommand{\arraystretch}{1.4}
\setlength\tabcolsep{0.3em} {
\begin{tabular}{cc|cccccc}
\hline
 \multicolumn{2}{c|}{Method} & Belis~\cite{Belis2024} & 2DOF & Sakh.~\cite{Sakhnenko2022} & RBF & Poly & Linear \\
\hline
\multirow{3}{*}{\rotatebox{90}{HAI}} & Tree & 2.71e-02 & 3.58e-03 & 1.90e-02 & 2.78e-03 & 2.11e-03 & 6.80e-04 \\
& NMF & 1.74e-04 & 1.94e-05 & 6.56e-05 & 1.01e-05 & 1.00e-05 & 5.02e-06 \\
& PCA & 5.46e-03 & 1.89e-03 & 3.96e-03 & 8.92e-04 & 7.81e-04 & 4.28e-04 \\
\hline
\multirow{3}{*}{\rotatebox{90}{SWaT}} & Tree & 2.31e-01 & 4.37e-02 & 2.10e-01 & 5.01e-02 & 5.62e-02 & 2.49e-02 \\
& NMF & 1.95e-01 & 1.07e-01 & 2.04e-01 & 1.24e-01 & 1.03e-01 & 6.64e-02 \\
& PCA & 2.10e-01 & 6.29e-02 & 1.81e-01 & 4.20e-02 & 4.66e-02 & 2.16e-02 \\
\hline
\multirow{3}{*}{\rotatebox{90}{WADI}} & Tree & 1.39e-01 & 2.64e-01 & 1.80e-01 & 1.88e-01 & 1.91e-01 & 1.00e-01 \\
& NMF & 9.65e-02 & 5.91e-02 & 1.05e-01 & 9.08e-02 & 7.89e-02 & 6.19e-02 \\
& PCA & 1.10e-01 & 1.13e-01 & 1.50e-01 & 6.92e-02 & 6.90e-02 & 3.39e-02 \\
\hline
\end{tabular}
}
\end{table}

The expectation is that if there is a quantifiable advantage to quantum kernels, there should be a significant difference between the classical and quantum KTAs. Interestingly, the target alignments reflect just that; there is a significant positive difference of almost 4.8\% KTA. 

In HAI, Belis Feature Map maintains an order of magnitude higher kernel alignment in all pre-processing methods, followed by Sakhnenko and the simple 2DoF. The classical kernels have distinctly lower KTAs, and the classification results described before reflect this. The HAI dataset was very non-linear and complex, which can possibly explain this difference.

For SWaT, the comparison is not as clear. The Belis and Sakhnenko kernels had the best target alignment out of all, but the polynomial and linear kernel functions were reportedly better than the simple 2DoF model. However, this allows us to see that even some low-performing models may have higher geometric distance than seemingly better classifiers. This further emphasizes that kernel alignment is important, but not the only aspect of a good-performing model.

Finally, WADI represents an equal comparison of kernel alignment between quantum feature maps and classical functions and how it can affect classification. Quantum performed better with the Decision Tree feature selection, but the classical achieved similar results in NMF and PCA. Looking at the WADI section in Table \ref{tbl:quant_advantage}, a similar trend appears in the kernel target alignment. In ``Tree", 2DoF achieves much better classification and alignment results; however, in all other pre-processing methods, the classical and quantum were very much alike in classification and alignment.

On average, there presents a 0.048 increase for quantum kernels in kernel alignment across all the datasets. This presents a 91.023\% percent change over classical, demonstrating a higher kernel alignment for quantum over classical kernel functions.

\section{Discussion, Conclusion, and Future Work} \label{sec:Conclusion}
\subsection{Discussion}
Quantum has shown superiority in the classification of all three datasets, demonstrating a fairly clear quantitative advantage to quantum anomaly detection methods in our experiments. Our quantum experiments demonstrated a 13.3\% increase in F1 classification and 91.023\% stronger kernel-target alignment than classical methods. However, classification and KTA performance are hardly the only deciding factors to safe and effective real-world implementation. A summarization of the best kernel configurations and feature counts for all three datasets are provided in Table \ref{tbl:summary}. We can see that the best two feature maps were the simplest: Belis et al.'s kernel and the simple U2-gate ``2DoF" kernel, both presenting over 0.86 F1 score over all three datasets. This shows that the simple models thrive better as they avoid overfitting models and still provide quantum efficiency to the kernel calculations. Additionally, we can see that tree selection preprocessing to 12 and 16 features (6 and 8 qubits) presented the best results, likely due to feature importance playing a role.

\subsection{Conclusion and Future Work}
In conclusion, we have effectively investigated the use of Quantum Hybrid Support Vector Machines in the CPS anomaly detection domain and shown a clear quantitative advantage to using them over their classical counterparts for ICS data. With a consistently significant improvement in both classification and alignment in all three real-world datasets, our quantum kernels demonstrated strong potential for applicability, even in a noisy environment such as NISQ computing.
As shown in this work, deep parametrization and experimentation is essential for QAD. For example, if a feature map is too simple, the fundamental concepts are extremely simplified, resulting in possibly little to no quantum advantage and easy classical reproducibility~\cite{Havlicek2019}. Similarly, quantum fidelity error can increase possibly exponentially with qubit count, a trend analogous to the ``curse of dimensionality"~\cite{Shaydulin2022}. This can cause a model to quickly overfit to the highly-exponential system and destroy classification performance. These concerns vary across application domains, requiring strong consideration.

% Table of summarized results
\begin{table}[tbp]
\centering
\caption{Summary of best classification metrics for quantum kernels on ICS/CPS data.}
\label{tbl:summary}
\renewcommand{\arraystretch}{1.4}
\setlength\tabcolsep{0.3em} {
\begin{tabular}{c|ccc|ccc}
\hline
 & \multicolumn{3}{c|}{Best Configuration:} & \multicolumn{3}{c}{Results:} \\
 Dataset & Kernel & Feature & Method & Precision & Recall & F1 Score \\
\hline
SWaT & 2DoF & 12 & Tree & 0.95 & 0.95 & 0.95 \\
HAI & Belis \cite{Belis2024} & 16 & Tree & 0.88 & 0.87 & 0.86 \\
WADI & 2DoF & 16 & Tree & 0.90 & 0.88 & 0.88 \\
\hline
\end{tabular}
}
\end{table}

Additionally, longer circuits are often more noisy and can easily drop classification performance, as shown in previous QML literature. Thus, feature maps with more gate operations and complexity create more noise. While noise-mitigated algorithms are beginning to surface in quantum-cloud computing~\cite{Google2024}, the effect of noise is still a confounding variable that must be considered and alleviated. 

Furthermore, NISQ computing is highly limited in availability and practicality resulting in lengthy queues for resources and very limited computing time~\cite{Asiani2021}. This is currently infeasible for large datasets as the number of total shots would be upper bounded by $O(\lvert x_{input} \rvert \times \lvert x_{train} \rvert \times shots)$. Thus, past literature usually uses less than 500 data point samples for QML in their experimentation. For kernel applications, large feature maps with many hyper-parametrized gates still remain a theory with no current feasible application.

There are many practicality and availability problems in modern quantum computing that greatly restrict the immediate commercialization of a responsive quantum-enabled AD system. However, with the continuing trend of rapid technical advances in Quantum~\cite{Google2024,lotstedt2024}, the results presented here present a clear potential in the advantages of quantum computing. 

This work pulls together quantum machine learning with real-world applications through thorough study. Future work can build upon this investigation by performing deeper analyses with future quantum computing resources, kernel algorithms, and feature maps. With better computing resources, the highly exhaustive number of quantum shots for kernels can be optimized, more qubits can enable more features, and intricate, low-noise feature maps can be better implemented. On a related note, our work only features quantum simulations of real computer noise models. While these noise models are highly realistic to current IBM systems, this study must be extended to actual shots on real hardware, including alternative quantum technologies like ion-trap machines.

As the realm of quantum kernel computing only continues to grow in applicability and accessibility in the real world, research should continue to determine the true impact the quantum kernel domain can have on the cyberphysical computing space surrounding us everyday.

\section*{Acknowledgment}

This research used resources from the Oak Ridge Leadership Computing Facility, which is a DOE Office of Science User Facility supported under Contract DE-AC05-00OR22725. This material is based upon work supported by the National Science Foundation Graduate Research Fellowship under Grant No. 1938092.

\ifCLASSOPTIONcaptionsoff
  \newpage
\fi

\bibliographystyle{IEEEtran}
\bibliography{IEEEabrv,reference}

\vspace{-.5in}
\begin{IEEEbiographynophoto}{Tyler Cultice}{\,} is pursuing his PhD with the  Department of Electrical Engineering and Computer Science, University of Tennessee, Knoxville, TN 37996, USA. Contact him at tcultice@vols.utk.edu.
\end{IEEEbiographynophoto}
\vspace{-.5in}
\begin{IEEEbiographynophoto}{Md. Saif Hassan Onim}{\,} is pursuing his PhD with the  Department of Electrical Engineering and Computer Science, University of Tennessee, Knoxville, TN 37996, USA. Contact him at monim@vols.utk.edu.
\end{IEEEbiographynophoto}

\vspace{-.5in}
\begin{IEEEbiographynophoto}{Annarita Giani}{\,} is a Senior Scientist with the GE Vernova Advanced Research Center, Niskayuna, NY, USA.
\end{IEEEbiographynophoto}\balance

\vspace{-.5in}
\begin{IEEEbiographynophoto}{Himanshu Thapliyal}{\,} is an Associate Professor with the  Department of Electrical Engineering and Computer Science, University of Tennessee, Knoxville, TN 37996, USA. Contact him at hthapliyal@utk.edu.
\end{IEEEbiographynophoto}\balance

\end{document}